\documentclass[structabstract]{aa}  
\usepackage{graphicx}
\usepackage{mathtools}
\usepackage{txfonts}
\newcommand{\teff}{\ifmmode T_{\rm eff} \else $T_{\mathrm{eff}}$\fi}
\newcommand{\logg}{\ifmmode \log g \else $\log g$\fi}
\newcommand{\lL}{\ifmmode \log \frac{L}{L_{\odot}} \else $\log \frac{L}{L_{\odot}}$\fi}

\newcommand{\vsini}{\ifmmode v \sin i \else $v \sin i$\fi}

\newcommand{\kms}{km~s$^{-1}$}
\newcommand{\msun}{\ifmmode M_{\odot} \else $M_{\odot}$\fi}
\newcommand{\zsun}{\ifmmode Z_{\odot} \else $Z_{\odot}$\fi}
\newcommand{\lsun}{\ifmmode L_{\odot} \else $L_{\odot}$\fi}
\newcommand{\rsun}{\ifmmode R_{\odot} \else $R_{\odot}$\fi}
\newcommand{\qh}{\ifmmode Q_{\rm H} \else $Q_{\rm H}$\fi}
\newcommand{\qhei}{\ifmmode Q_{\ion{He}{i}} \else $Q_{\ion{He}{i}}$\fi}

\begin{document}
   \title{The Tarantula Massive Binary Monitoring.}

   \subtitle{IV. Double-lined photometric binaries}

   \author{L. Mahy\inst{1}
         \and
          L. A. Almeida\inst{2,3}
           \and
          H. Sana\inst{1}
          \and
          J. S. Clark\inst{4}
          \and
          A. de Koter\inst{5,1}
          \and
          S. E. de Mink\inst{6,5}
           \and
          C. J. Evans\inst{7}
         \and
          N. J. Grin\inst{8}
          \and
          N. Langer\inst{8}
          \and
          A. F. J. Moffat\inst{9}
          \and
          F. R. N. Schneider\inst{10,11}
          \and
          T. Shenar\inst{1}
          \and
          F. Tramper\inst{12}
           }

   \offprints{L. Mahy}

   \institute{Instituut voor Sterrenkunde, KU Leuven, Celestijnlaan 200D, Bus 2401, B-3001 Leuven, Belgium\\
   \email{laurent.mahy@kuleuven.be}
   \and
   Departamento de F\'isica, Universidade do Estado do Rio Grande do Norte, Mossor\'o, RN, Brazil
   \and 
   Departamento de F\'isica, Universidade Federal do Rio Grande do Norte, UFRN, CP 1641, Natal, RN, 59072-970, Brazil
   \and
   School of Physical Sciences, The Open University, Walton Hall, Milton Keynes, MK7 6AA, UK
   \and
   Astronomical Institute Anton Pannekoek, Amsterdam University, Science Park 904, 1098 XH Amsterdam, The Netherlands
   \and
   Center for Astrophysics, Harvard \& Smithsonian, 60 Garden Street, Cambridge, MA 02138, USA
   \and
   UK Astronomy Technology Centre, Royal Observatory Edinburgh, Blackford Hill, Edinburgh EH9 3HJ, United Kingdom
   \and
   Argelander-Institut f{\"u}r Astronomie der Universit{\"a}t Bonn, Auf dem H{\"u}gel 71, 53121 Bonn, Germany
  \and
   D{\'e}partement de physique, Universit{\'e} de Montr{\'e}al, and Centre de Recherche en Astrophysique de Qu{\'e}bec, CP 6128, Succ. C-V, Montr{\'e}al, QC, H3C 3J7, Canada
   \and
  Zentrum f\"{u}r Astronomie der Universit\"{a}t Heidelberg, Astronomisches Rechen-Institut, M\"{o}nchhofstr. 12-14, 69120 Heidelberg, Germany
  \and
   Heidelberger Institut f\"{u}r Theoretische Studien, Schloss-Wolfsbrunnenweg 35, 69118 Heidelberg, Germany
   \and
   IAASARS, National Observatory of Athens, 15326 Penteli, Greece
   }
   
   \date{Received ...; accepted ...}

 
  \abstract
   {A high fraction of massive stars are found to be binaries but only a few of them are reported as photometrically variable. By studying the populations of double-lined spectroscopic binaries in the 30 Doradus region, we found a subset of them that have photometry from the OGLE project and that display variations in their light curves related to orbital motions.}
   {The goal of this study is to determine the dynamical masses and radii of the 26 binary components in order to investigate the mass-discrepancy problem and to provide an empirical mass-luminosity relation for the Large Magellanic Cloud.}
   {We use the PHOEBE programme to perform a systematic analysis of the OGLE $V$ and $I$ light curves obtained for 13 binary systems in the 30 Doradus region. We adopt the effective temperatures, and orbital parameters derived previously to obtain the inclinations of the systems and the parameters of the individual components.  }
   {Three systems display eclipses in their light curves, while the others only display ellipsoidal variations. We classify two systems as over-contact, five as semi-detached, and four as detached. The two remaining systems have uncertain configurations due to large uncertainties on their inclinations. The fact that systems display ellipsoidal variations has a significant impact on the inclination errors. From the dynamical masses, luminosities, and radii, we provide LMC-based empirical mass-luminosity and mass-radius relations, and we compare them to other relations given for the Galaxy, the Large Magellanic Cloud (LMC), and the Small Magellanic Cloud (SMC). These relations differ for different mass ranges, but do not seem to depend on the metallicity regimes. We also compare the dynamical, spectroscopic, and evolutionary masses of the stars in our sample. While the dynamical and spectroscopic masses agree with each other, the evolutionary masses are systematically higher, at least for stars in semi-detached systems. This suggests that the mass discrepancy can be partly explained by past or ongoing interactions between the stars.}
   {}

   \keywords{Stars: early-type - Stars: binaries: photometry - Stars: fundamental parameters - Open clusters and associations: individual: 30\,Doradus}

   \maketitle


\section{Introduction}
\label{intro}

Massive stars play a crucial role in the ecology of galaxies, providing mechanical and radiative feedback on their host environments. Their fate is primarily governed by their mass, rotation, and metallicity. They end their lives as core-collapse supernovae, and can produce compact remnants such as neutron stars or black holes. Moreover, massive stars are commonly found in pairs or multiple systems \citep[see e.g.][and references therein]{sana12, sana13, sana14}. The presence of a nearby companion significantly alters the way that the stars evolve \citep{pacynski71, pod92, demink13}. Interactions during the main sequence are expected in the closest binaries, whilst in longer period systems ($P_{\rm ini} > 6$~days) the interactions are expected to occur after completion of the main sequence evolution such that they evolve most of their life as single stars.

The ground-breaking detection of gravitational waves by Advanced-LIGO detectors \citep{abbott16} provided observational evidence that binary black hole systems exist.This discovery opened a new decisive phase in the study of massive stars, and emphasised that our understanding of the evolution of massive stars is not yet complete. It also pointed out the need for accurate stellar parameters of massive stars to compare with binary evolution models. 

In this context, eclipsing binary systems are one of the primary sources used to derive the fundamental properties of stars (masses, radii, and luminosities). The high accuracy that can be reached on the stellar properties has greatly contributed to improving stellar evolution theory \citep[e.g.][]{harries03,hilditch05}. Before the {\em Gaia} era, eclipsing binaries were also used to determine accurate distances of the clusters or galaxies in which they were embedded \citep[see e.g.][]{Vilardell10,north10,Kourniotis15}. Moreover, they have played important roles in probing the possible interactions between these two components that totally modify the way that they evolve \citep{nelson01,demink17}. 

In the past decade, the massive star population of the Tarantula nebula (also known as 30 Doradus) was the focus of a large multi-epoch optical spectroscopic campaign (the VLT-FLAMES Tarantula Survey VFTS; \citealt{evans11}, PI: Evans, 182.D-0222). Subsequent observations were then obtained by the Tarantula Massive Binary Monitoring programme  (TMBM, PI: Sana, 090.D-0323 and 092.D-0136), which was designed to characterise the properties of the binary systems identified in this region. Among the massive binaries observed by TMBM, 51 were reported as single-lined spectroscopic binaries (SB1) and 31 as double-lined spectroscopic binaries (SB2). The properties of the orbits were provided by \citet[][hereafter Paper I,]{almeida17} whilst the stellar parameters ($L$, \teff, \logg, \vsini) and surface abundances (He, C, and N) of all the components in the SB2 systems were given by Mahy et al. (2020, in press, hereafter Paper III). Among these objects, 13 show clear photometric variations in their light curves from Phase IV of the Optical Gravitational Lensing Experiment (OGLE, \citealt{udalski15}).
 
The present paper focuses on the analysis of the light curves for these 13 massive binary systems located in 30\,Dor. Our goal is to determine the dynamical masses and the radii of their components. Here we use them to derive mass-luminosity and mass-radius relations at LMC (Large Magellanic Cloud) metallicity and to investigate the mass-discrepancy problem \citep{herrero92}. The paper is organised as follows. In Sect.\,\ref{obs} we describe the observations and the data analysis procedure. The results for the 13 individual binary systems are provided in Section\,\ref{results} and are discussed in Section\,\ref{discussion}. Finally, we summarise and draw our conclusions in Section\,\ref{conclusion}.


\section{Observations and data analysis}
\label{obs}

Among the 31 SB2 binary systems analysed in Papers I and II, 13 were identified by the OGLE project as showing photometric variations. OGLE $V$- and $I$-band photometry were obtained from phase IV of the project \citep{udalski15}. The data were obtained between 2010 March and 2014 March for these 13 systems. 

The OGLE light curves were analysed with the software programme PHOEBE \citep[PHysics Of Eclipsing BinariEs, v0.31a,][]{prsa05}. PHOEBE models the multi-bandpass light curves and the radial velocity (RV) curves of the object at the same time. This software is based on the code of Wilson and Devinney \citep{wilson71}, and uses the Nelder and Mead Simplex fitting method \citep{nelder65} to adjust all the input parameters and to find the best fits to the light curves. For the Nelder and Mead Simplex, the maximum number of iterations is fixed to 200, and the aimed relative precision to $10^{-3}$.

Given that all considered components have radiative envelopes, we fix the gravity darkening ($g_i$) exponent to unity for all components \citep{vonzeipel24}, and also the surface albedos ($A_i = 1.0$). For the limb darkening, we use the square root law, better suited for hotter stars \citep{diaz95}. The option of a reflection effect is enabled and set to two reflections. All the orbits are taken to be circular. A $\chi^2$ minimisation procedure computed between the synthesised light curves and the observed data is used to evaluate the quality of the solutions.

To complete our dataset, we use the orbital parameters given in Paper III(the orbital periods, projected semi-major axes, mass ratios and HJD$_0$) and the RVs that were derived in Paper I. We keep the effective temperatures of the stars (derived after spectral disentangling and atmosphere fitting in Paper III) fixed for our analysis, and only consider the Roche lobe filling factors (and thus the configurations of the systems) and the inclinations as free parameters. The uncertainties of the effective temperatures are neglected. As specified in Paper III, the current most massive stars of the systems are defined as primaries. 

\begin{sidewaystable*}
\scriptsize
\caption{Masses, radii, and inclinations determined for all the binary systems and components in our sample. Errors represent 1-$\sigma$.}             
\label{tab:parameters}      
\centering                          
\begin{tabular}{l r r r r r r r r r r r r r r r r}        
\hline\hline                 
Star  & $P_{\rm orb}$ & $M_P/M_S$ & $\log(L/L_{\odot})$ & \teff \tablefootmark{a} & $R^{\mathrm{mean}}$ & $R_{\mathrm{RL}}$ & $R/R_{\mathrm{RL}}$& $M_{\rm dyn}$ &$M_{\rm spec}$ &$M_{\rm evol}$  & $M_{\rm dyn}/M_{\rm spec}$& $M_{\rm spec}/M_{\rm evol}$& $M_{\rm dyn}/M_{\rm evol}$ & $i$ & $a$ & rms\\
VFTS         &[d] &   &   & [kK]          &      [\rsun] & [\rsun] &  & [\msun]& [\msun]& [\msun]  &   &  &   &[$\degr$] &      [\rsun]  & [mag]\\
\hline       
\multicolumn{12}{c}{Detached binary} \\
\hline  

500P  & $2.875370$  &  $1.05 \pm 0.01$ & $5.29_{-0.06}^{+0.07}$  & $40.7_{- 0.6}^{+ 0.7}$ & $ 8.9_{- 0.6}^{+ 0.6}$ & $11.9_{- 0.3}^{+ 0.3}$ & $0.75_{-0.05}^{+0.05}$ & $25.1_{- 1.5}^{+ 1.6}$ &  $31.1_{-5.8}^{+4.7}$ & $32.8_{-1.3}^{+1.5}$  & $0.81_{-0.16}^{+0.13}$ & $0.95_{-0.18}^{+0.15}$ & $0.77_{-0.05}^{+0.06}$ & $61.2_{- 2.0}^{+ 2.1}$ & $31.1_{-0.7}^{+0.6}$ & 0.006\\ [4pt]

500S  &           &           & $5.21_{-0.07}^{+0.08}$  & $39.4_{- 0.8}^{+ 0.8}$ & $ 8.7_{- 0.6}^{+ 0.8}$ & $11.6_{- 0.2}^{+ 0.3}$ & $0.75_{-0.05}^{+0.07}$  & $23.8_{- 1.5}^{+ 1.5}$ &  $29.3_{-5.0}^{+5.0}$ & $29.8_{-1.2}^{+1.6}$ & $0.81_{-0.15}^{+0.15}$ & $0.98_{-0.17}^{+0.18}$ & $0.80_{-0.06}^{+0.07}$  &   &   &  \\ [4pt]

543P  & $1.383990$  &  $1.13\pm 0.04$ & $4.61_{-0.14}^{+0.09}$  & $33.2_{- 1.4}^{+ 1.4}$ & $ 6.1_{- 0.8}^{+ 0.2}$ & $ 7.1_{- 0.5}^{+ 0.6}$ & $0.86_{-0.13}^{+0.08}$ & $22.5_{- 4.4}^{+ 6.0}$ &  $23.4_{-11.6}^{+10.8}$ & $17.4_{-1.4}^{+1.5}$ & $0.96_{-0.51}^{+0.51}$ & $1.35_{-0.68}^{+0.63}$ & $1.29_{-0.27}^{+0.36}$  & $37.2_{- 2.7}^{+ 3.8}$ & $18.2_{-1.6}^{+1.2}$ & 0.007\\ [4pt]

543S  &           &           & $4.51_{-0.15}^{+0.06}$  & $32.4_{- 0.7}^{+ 1.0}$ & $ 5.7_{- 0.9}^{+ 0.2}$ & $ 6.7_{- 0.4}^{+ 0.6}$ & $0.85_{-0.14}^{+0.08}$  & $20.0_{- 3.9}^{+ 5.4}$ &  $21.2_{-7.2}^{+4.5}$ & $16.0_{-1.0}^{+1.0}$ & $0.94_{-0.37}^{+0.32}$ & $1.33_{-0.46}^{+0.29}$ & $1.25_{-0.26}^{+0.35}$ &    &  & \\ [4pt]

642P  & $1.726820$  &  $1.55\pm 0.02$ & $5.08_{-0.13}^{+0.18}$  & $40.7_{- 0.6}^{+ 1.6}$ & $ 7.0_{- 1.0}^{+ 1.4}$  & $ 9.3_{- 1.2}^{+ 1.2}$ & $0.75_{-0.14}^{+0.18}$  & $29.8_{-11.7}^{+11.7}$ &  $24.9_{-6.6}^{+9.2}$ & $29.0_{-1.9}^{+2.0}$ & $1.20_{-0.57}^{+0.65}$ & $0.86_{-0.24}^{+0.32}$ & $1.03_{-0.41}^{+0.41}$ & $33.7_{- 5.0}^{+ 5.0}$ & $22.2_{-2.9}^{+2.9}$& 0.006\\ [4pt]

642S  &           &           & $4.68_{-0.19}^{+0.19}$  & $34.8_{- 1.9}^{+ 2.5}$ & $ 6.0_{- 1.1}^{+ 1.1}$ & $ 7.5_{- 1.0}^{+ 1.0}$ & $0.80_{-0.18}^{+0.18}$ & $19.2_{- 7.6}^{+ 7.6}$ &  $17.8_{-10.9}^{+11.4}$ & $18.6_{-2.0}^{+2.2}$ & $1.08_{-0.79}^{+0.82}$ & $0.95_{-0.60}^{+0.62}$ & $1.03_{-0.42}^{+0.43}$ &    &   &\\ [4pt]

661P  & $1.266430$  &  $1.41\pm 0.02$ & $4.96_{-0.02}^{+0.04}$  & $38.4_{- 0.4}^{+ 0.9}$ & $ 6.8_{- 0.0}^{+ 0.0}$ & $ 7.5_{- 0.1}^{+ 0.1}$ & $0.91_{-0.02}^{+0.02}$ & $27.3_{- 1.0}^{+ 0.9}$ &  $25.8_{-2.4}^{+4.2}$ & $26.0_{-1.0}^{+1.3}$ & $1.06_{-0.10}^{+0.17}$ & $0.99_{-0.10}^{+0.17}$ & $1.05_{-0.06}^{+0.06}$ & $64.5_{- 0.6}^{+ 0.2}$ & $17.7_{-0.2}^{+0.2}$& 0.007\\ [4pt]

661S  &           &           & $4.48_{-0.03}^{+0.08}$  & $31.8_{- 0.6}^{+ 1.4}$ & $ 5.7_{- 0.0}^{+ 0.0}$   & $ 6.1_{- 0.1}^{+ 0.1}$ & $0.93_{-0.02}^{+0.02}$ & $19.4_{- 0.7}^{+ 0.6}$ &  $17.5_{-3.6}^{+6.0}$ & $16.6_{-0.8}^{+0.9}$& $1.11_{-0.23}^{+0.38}$ & $1.06_{-0.22}^{+0.37}$ & $1.17_{-0.07}^{+0.07}$  &    &  &  \\ [4pt]

\hline       
\multicolumn{12}{c}{Semi-detached binary} \\
\hline  
061P  & $2.333440$  &  $1.88 \pm 0.11$ & $4.77_{-0.05}^{+0.04}$  & $33.5_{- 0.9}^{+ 0.6}$ & $ 7.2_{- 0.2}^{+ 0.2}$ & $ 9.4_{- 0.3}^{+ 0.3}$ & $0.77_{-0.03}^{+0.03}$  & $16.3_{- 1.4}^{+ 1.4}$ & $17.8_{-4.5}^{+3.7}$ & $19.0_{-0.7}^{+0.6}$ & $0.92_{-0.25}^{+0.21}$ & $0.94_{-0.24}^{+0.20}$ & $0.86_{-0.08}^{+0.08}$ & $69.1_{- 0.9}^{+ 0.9}$& $21.6_{-0.7}^{+0.7}$ & 0.013 \\ [4pt]

061S  &           &           & $4.75_{-0.05}^{+0.05}$  & $32.9_{- 0.6}^{+ 0.7}$ & $ 7.3_{- 0.3}^{+ 0.3}$  & $ 7.0_{- 0.3}^{+ 0.3}$ & $1.04_{-0.06}^{+0.06}$ & $ 8.7_{- 0.6}^{+ 0.6}$ &  $9.2_{-2.1}^{+2.7}$ & $22.2_{-1.6}^{+3.6}$ & $0.95_{-0.22}^{+0.28}$ & $0.41_{-0.10}^{+0.14}$ & $0.39_{-0.04}^{+0.07}$ &     &  &  \\ [4pt]

094P  & $2.256330$  &  $1.08 \pm 0.11$ & $5.51_{-0.13}^{+0.09}$  & $41.9_{- 0.4}^{+ 0.4}$ & $10.9_{- 1.7}^{+ 1.1}$ & $10.9_{- 1.2}^{+ 1.3}$ & $1.00_{-0.19}^{+0.16}$  & $30.5_{-10.6}^{+11.6}$ &  $31.4_{-9.8}^{+6.5}$ & $43.8_{-3.1}^{+3.9}$ & $0.97_{-0.45}^{+0.42}$ & $0.72_{-0.23}^{+0.16}$ & $0.70_{-0.25}^{+0.27}$ & $28.1_{- 3.0}^{+ 3.4}$ & $28.1_{-3.4}^{+3.0}$& 0.010 \\ [4pt]

094S  &           &           & $5.19_{-0.12}^{+0.14}$  & $40.1_{- 1.4}^{+ 1.4}$ & $ 8.1_{- 1.1}^{+ 1.2}$ & $10.5_{- 1.2}^{+ 1.3}$ & $0.77_{-0.14}^{+0.15}$  & $28.2_{- 9.8}^{+10.7}$ &  $27.7_{-9.7}^{+11.0}$ & $29.8_{-2.1}^{+2.7}$ & $1.02_{-0.50}^{+0.56}$ & $0.93_{-0.33}^{+0.38}$ & $0.95_{-0.34}^{+0.37}$ &    &   & \\ [4pt]

176P  & $1.777590$  &  $1.62 \pm 0.05$ & $5.24_{-0.02}^{+0.02}$  & $38.3_{- 0.3}^{+ 0.3}$ & $ 9.4_{- 0.1}^{+ 0.1}$   & $ 9.3_{- 0.2}^{+ 0.2}$ & $1.01_{-0.02}^{+0.02}$  & $28.3_{- 1.5}^{+ 1.5}$ &  $24.6_{-1.3}^{+1.3}$ & $32.2_{-1.4}^{+1.2}$ & $1.15_{-0.08}^{+0.08}$ & $0.77_{-0.05}^{+0.05}$ & $0.88_{-0.06}^{+0.06}$ & $89.7_{- 0.3}^{+ 0.3}$ & $22.1_{-0.3}^{+0.3}$& 0.012 \\ [4pt]

176S  &           &           & $4.15_{-0.12}^{+0.09}$  & $28.5_{- 1.8}^{+ 1.4}$ & $ 4.9_{- 0.2}^{+ 0.2}$ & $ 7.5_{- 0.1}^{+ 0.1}$ & $0.65_{-0.03}^{+0.03}$  & $17.5_{- 0.9}^{+ 0.9}$ &  $16.3_{-2.1}^{+5.0}$ & $12.2_{-0.8}^{+1.0}$ & $1.07_{-0.15}^{+0.33}$ & $1.34_{-0.20}^{+0.42}$ & $1.43_{-0.12}^{+0.14}$ &   &   & \\ [4pt]

450P  & $6.892330$  &  $1.04 \pm 0.08$ & $5.30_{-0.10}^{+0.21}$  & $33.8_{- 2.3}^{+ 1.7}$ & $13.0_{- 0.6}^{+ 3.0}$  & $22.4_{- 0.9}^{+ 0.9}$ & $0.58_{-0.04}^{+0.14}$  & $29.0_{- 4.0}^{+ 4.1}$ &  $36.1_{-11.7}^{+10.4}$ & $25.0_{-1.3}^{+4.0}$& $0.80_{-0.28}^{+0.26}$ & $1.44_{-0.47}^{+0.47}$ & $1.16_{-0.17}^{+0.25}$   & $63.5_{- 1.2}^{+ 1.7}$ & $58.5_{-2.3}^{+2.3}$ & 0.023 \\ [4pt]

450S  &           &           & $5.46_{-0.04}^{+0.04}$  & $28.3_{- 0.2}^{+ 0.3}$ & $22.2_{- 0.4}^{+ 0.3}$  & $22.0_{- 0.9}^{+ 0.9}$ & $1.01_{-0.05}^{+0.04}$  & $27.8_{- 3.8}^{+ 3.9}$ &  $21.1_{-2.1}^{+2.4}$ & $24.6_{-2.0}^{+2.0}$ & $1.32_{-0.22}^{+0.24}$ & $0.86_{-0.11}^{+0.12}$ & $1.13_{-0.18}^{+0.18}$  &   &  & \\ [4pt] 

652P  & $8.589090$  &  $2.77 \pm 0.36$ & $5.35_{-0.04}^{+0.06}$  & $32.1_{- 0.7}^{+ 0.9}$ & $15.4_{- 0.3}^{+ 0.7}$  & $24.1_{- 2.1}^{+ 1.9}$ & $0.64_{-0.06}^{+0.06}$ & $18.1_{- 4.2}^{+ 3.6}$ &  $18.0_{-1.8}^{+3.2}$ & $29.6_{-2.5}^{+2.3}$ & $1.01_{-0.25}^{+0.27}$ & $0.61_{-0.08}^{+0.12}$ & $0.61_{-0.15}^{+0.13}$  & $63.7_{- 4.8}^{+ 0.9}$ & $51.3_{-3.6}^{+4.2}$& 0.040\\ [4pt]

652S  &           &           & $4.92_{-0.07}^{+0.06}$  & $23.9_{- 0.3}^{+ 0.5}$ & $16.8_{- 0.2}^{+ 0.7}$ & $15.1_{- 1.4}^{+ 1.3}$ & $1.11_{-0.10}^{+0.11}$  & $ 6.5_{- 1.1}^{+ 0.8}$ &  $6.6_{-2.0}^{+2.0}$ & $24.8_{-3.9}^{+11.5}$ & $0.98_{-0.34}^{+0.32}$ & $0.27_{-0.09}^{+0.15}$ & $0.26_{-0.06}^{+0.13}$ &    &  & \\ [4pt]

\hline       
\multicolumn{12}{c}{Contact binary} \\
\hline  
066P  & $1.141160$  &  $1.91 \pm 0.05$ & $4.54_{-0.13}^{+0.09}$  & $32.8_{- 1.0}^{+ 1.7}$ & $ 5.8_{- 0.8}^{+ 0.5}$ & $ 5.4_{- 0.8}^{+ 1.0}$ & $1.07_{-0.22}^{+0.22}$ & $13.0_{- 5.0}^{+ 7.0}$ &  $14.6_{-4.3}^{+5.2}$ & $16.6_{-0.9}^{+1.1}$ & $0.89_{-0.43}^{+0.58}$ & $0.88_{-0.26}^{+0.32}$ & $0.78_{-0.30}^{+0.43}$ & $17.5_{- 2.5}^{+ 3.2}$ & $12.3_{-1.7}^{+2.2}$ & 0.010\\ [4pt]

066S  &           &           & $4.08_{-0.18}^{+0.10}$  & $29.0_{- 1.2}^{+ 1.0}$ & $ 4.4_{- 0.8}^{+ 0.4}$ &$ 4.0_{- 0.6}^{+ 0.7}$ & $1.10_{-0.26}^{+0.22}$ & $ 6.6_{- 2.8}^{+ 3.5}$ &$ 6.9_{-2.9}^{+3.0} $ & $ 12.0_{-0.8}^{+0.7} $ & $0.95_{-0.57}^{+0.65}$ & $0.58_{-0.24}^{+0.25}$ & $0.55_{-0.24}^{+0.29}$   &   &  & \\ [4pt]

352P  & $1.124140$  &  $1.02 \pm 0.02$ & $5.10_{-0.04}^{+0.02}$  & $41.6_{- 1.0}^{+ 0.4}$ & $ 6.8_{- 0.2}^{+ 0.1}$ & $ 6.4_{- 0.1}^{+ 0.1}$ & $1.06_{-0.04}^{+0.02}$  & $25.6_{- 1.4}^{+ 1.7}$ &  $21.5_{-1.9}^{+1.1}$ & $22.8_{-1.0}^{+1.0}$ & $1.19_{-0.12}^{+0.10}$ & $0.94_{-0.09}^{+0.06}$ & $1.12_{-0.08}^{+0.09}$ & $53.6_{- 0.9}^{+ 1.3}$ & $16.8_{-0.3}^{+0.3}$& 0.009\\ [4pt]

352S  &           &           & $5.05_{-0.04}^{+0.02}$  & $40.6_{- 0.6}^{+ 0.4}$ & $ 6.8_{- 0.2}^{+ 0.1}$ & $ 6.4_{- 0.1}^{+ 0.1}$ & $1.06_{-0.04}^{+0.02}$  & $25.1_{- 1.4}^{+ 1.6}$ &  $22.2_{-1.9}^{+1.1}$ & $22.6_{-1.0}^{+2.2}$ & $1.13_{-0.12}^{+0.09}$ & $0.98_{-0.10}^{+0.11}$ & $1.11_{-0.08}^{+0.13}$ &   & &   \\ [4pt]

\hline       
\multicolumn{12}{c}{Uncertain configurations} \\
\hline  

217P  & $1.855340$  &  $1.20 \pm 0.01$ & $5.57_{-0.10}^{+0.13}$  & $45.0_{- 0.4}^{+ 0.6}$ & $10.1_{- 1.2}^{+ 1.5}$  & $11.1_{- 0.9}^{+ 0.9}$ & $0.91_{-0.13}^{+0.15}$   & $46.8_{-11.7}^{+11.7}$ &  $46.6_{-8.8}^{+11.2}$ & $44.8_{-2.8}^{+3.3}$& $1.00_{-0.31}^{+0.35}$ & $1.04_{-0.21}^{+0.26}$ & $1.04_{-0.27}^{+0.27}$  & $40.0_{- 4.0}^{+ 4.0}$ & $28.0_{-2.3}^{+2.3}$ & 0.007\\ [4pt]

217S  &           &           & $5.38_{-0.10}^{+0.13}$  & $41.8_{- 0.6}^{+ 1.7}$ & $ 9.4_{- 1.0}^{+ 1.4}$ & $10.2_{- 0.9}^{+ 0.9}$ & $0.92_{-0.13}^{+0.16}$  & $38.9_{- 9.7}^{+ 9.7}$ &  $38.7_{-8.0}^{+8.6}$ & $35.4_{-2.3}^{+2.7}$ & $1.00_{-0.32}^{+0.34}$ & $1.09_{-0.24}^{+0.26}$ & $1.10_{-0.28}^{+0.29}$  &    &  & \\ [4pt]

563P  & $1.217340$  &  $1.31 \pm 0.07$ & $4.63_{-0.10}^{+0.09}$  & $32.4_{- 0.9}^{+ 1.0}$ & $ 6.6_{- 0.7}^{+ 0.4}$ & $ 6.9_{- 0.5}^{+ 1.1}$ & $0.96_{-0.12}^{+0.16}$ & $26.2_{- 5.2}^{+11.9}$ &  $23.8_{-5.2}^{+8.0}$ & $17.6_{-1.1}^{+1.2}$ & $1.10_{-0.33}^{+0.62}$ & $1.35_{-0.31}^{+0.46}$ & $1.49_{-0.31}^{+0.68}$  & $29.7_{- 2.0}^{+ 4.9}$ & $17.2_{-2.6}^{+1.1}$ & 0.006\\ [4pt]

563S  &           &           & $4.52_{-0.12}^{+0.08}$  & $32.4_{- 0.9}^{+ 1.2}$ & $ 5.8_{- 0.6}^{+ 0.4}$  & $ 6.1_{- 0.4}^{+ 0.9}$ & $0.95_{-0.12}^{+0.15}$ & $20.0_{- 3.9}^{+ 9.1}$ &  $17.8_{-4.6}^{+4.0}$ & $16.4_{-1.1}^{+1.1}$ & $1.12_{-0.36}^{+0.57}$ & $1.09_{-0.29}^{+0.26}$ & $1.22_{-0.25}^{+0.56}$  &   &  &\\ [4pt]
\hline
\end{tabular}
\tablefoot{
\tablefoottext{a}{Effective temperatures are estimated with CMFGEN and their determination is  described in Paper III.}
}
\end{sidewaystable*}


\section{Results}
\label{results}

Based on the ratio of the radius of the components to that of their Roche lobe, we classify the 13 systems in four different categories: 
\begin{itemize}
\item detached when both components are well within their Roche lobe;
\item semi-detached when one component fills its Roche lobe;
\item contact/over-contact when both components fill their Roche lobes;
\item uncertain: when the uncertainties on the Roche lobe filling factors are too large to allow reliable classification of these systems (e.g. because of the low inclinations of the systems, and that only ellipsoidal variations are detected)
\end{itemize}
The inclinations ($i$), dynamical masses ($M_{\rm dyn}$), and radii ($R$) obtained by light-curve fitting are provided in Table\,\ref{tab:parameters}. We also provide the orbital periods ($P_{\rm orb}$) and mass ratios  ($M_P/M_S$) of the systems, as well as the effective temperatures (\teff), and spectroscopic ($M_{\rm spec}$) and evolutionary ($M_{\rm evol}$) masses of the components. We compute the luminosities ($L$) of the stars from their radius and their effective temperature (the solar luminosity value was taken from \citealt{mamajek15}). The radius of the Roche lobe ($R_{\mathrm{RL}}$) is computed from \citet{eggleton83}. As an example, Fig.\,\ref{fig:061} shows the light curves of VFTS\,061 and their best fits with PHOEBE. The other fits are shown in the Appendix.

The error bars given in Table\,\ref{tab:parameters} are established by exploring the parameter space. For this purpose we fix one parameter and allow the others to vary in order to reach the minimum of the $\chi^2$. The error bars are then determined for a variation in the $\chi^2$ corresponding to a 68.3\% confidence level ($1-\sigma$) and the appropriate number of degrees of freedom.

\subsection{Detached systems}

\subsubsection{VFTS\,500}
\label{result:VFTS500}

VFTS\,500 is classified as an O6\,V + O6.5 V binary system. It has an orbital period of 2.88 days. Its light curve (Fig.\,\ref{fig:500}) shows grazing eclipses and ellipsoidal variations. \citet{morrell14} reported an inclination of $64$\degr\ and individual masses of $20.5\pm 0.6\,\msun$ for the primary and $20.2 \pm 0.6\,\msun$ for the secondary. From effective temperatures of 40700\,K for the primary and 39400\,K for the secondary, we derive an inclination of $61.2_{-2.0}^{+2.1}$\degr. This value infers masses of $M_P = 25.1_{- 1.5}^{+ 1.6}\,\msun$ and $M_S = 23.8_{- 1.5}^{+ 1.5}\,\msun$. We compute radii of $R_P = 8.9_{-0.6}^{+0.6}\,\rsun$ and $R_S = 8.7_{-0.6}^{+0.8}\,\rsun$. The stars fill 42\% of their Roche lobes.

\subsubsection{VFTS\,543}
\label{result:VFTS543}

VFTS\,543 is composed of an O9.5\,V primary and an O9.7\,V secondary in a 1.38-day orbit. The light curves (Fig.\,\ref{fig:543}) display ellipsoidal variations, with $\Delta m = 0.06$ in the $V$ band and $\Delta m = 0.04$ in the $I$ band. We use as effective temperatures 33200\,K for the primary and 32400\,K for the secondary. We estimate an inclination of $37.2_{-2.7}^{+3.8}$\degr\ for the system. This value gives us masses of $M_P = 22.5_{- 4.4}^{+ 6.0}\,\msun$ and $M_S = 20.0_{- 3.9}^{+ 5.4}\,\msun$. The radii are estimated to $R_P = 6.1_{- 0.80}^{+ 0.2}\,\rsun$ and $R_S = 5.7_{- 0.9}^{+ 0.2}\,\rsun$. The stars fill about 60\% of their Roche lobes.

\subsubsection{VFTS\,642}
\label{result:VFTS642}

VFTS\,642 was classified as an O5.5\,V $+$ O9\,V binary system. The system has an orbital period of 1.73 days. Its light curve (Fig.\,\ref{fig:642}) displays ellipsoidal variations with a difference in magnitude of about $\Delta m = 0.03$ in the $V$ band and $\Delta m =0.07$ in the $I$ band. The inclination of the system is determined to be $33.7 \pm 5.0$\degr. From effective temperatures of 40700\,K for the primary and of 34800\,K for the secondary, we compute radii of $R_P = 7.0_{-1.0}^{+1.4}\,\rsun$ and $R_S = 6.0_{-1.1}^{+1.1}\,\rsun$. The primary fills 43\% of its Roche lobe and the secondary 51\%. Their masses are estimated to $M_P = 29.8 \pm  11.7\,\msun$ and $ M_S = 19.2 \pm 7.6\,\msun$. 

\subsubsection{VFTS\,661}
\label{result:VFTS661}

VFTS\,661 is an O6.5\,V + O9.7\,V binary system where the two components almost fill their Roche lobe with a filling factor of about 75\% for the primary and 82\% for the secondary. The system has an orbital period of 1.27 days. Its light curve (Fig.\,\ref{fig:661}) exhibits clear eclipses. The system is seen under an inclination of $64.5_{-0.6}^{+0.2}$\degr. We fix the effective temperature of the primary to 38400\,K and that of the secondary to 31800\,K. The masses are estimated to be $M_P = 27.3_{- 1.0}^{+ 0.9}\,\msun$ and $M_S = 19.4_{- 0.7}^{+ 0.6}\,\msun$. The mean radii are measured to $R_P = 6.8_{-0.01}^{+0.04}\,\rsun$ and $R_S = 5.7_{-0.01}^{+0.03}\,\rsun$. 

\subsection{Semi-detached systems}

\subsubsection{VFTS\,061}
\label{result:VFTS061}
\begin{figure*}[htbp]
  \centering
    \includegraphics[width=9cm, angle=0]{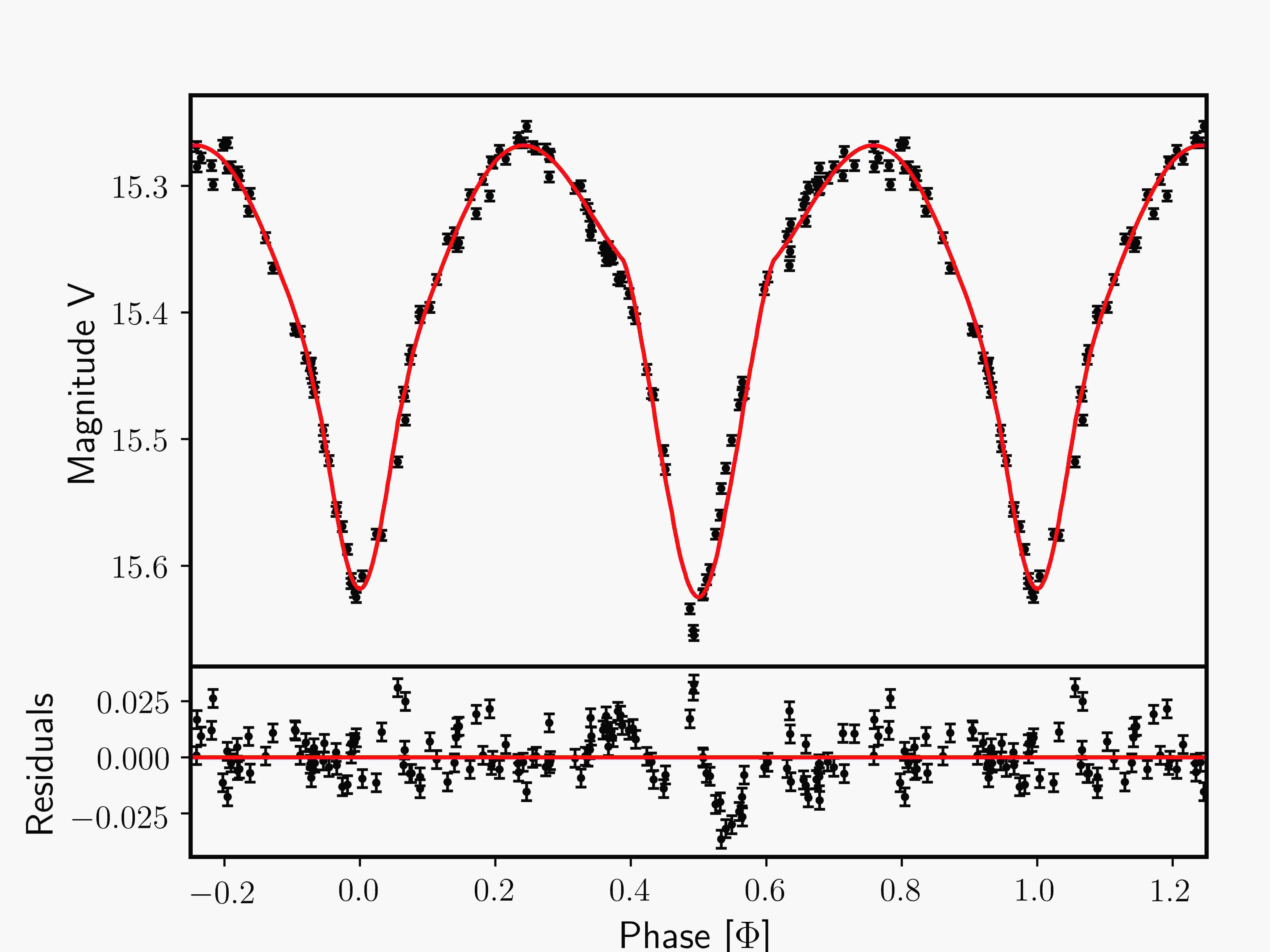}
    \includegraphics[width=9cm, angle=0]{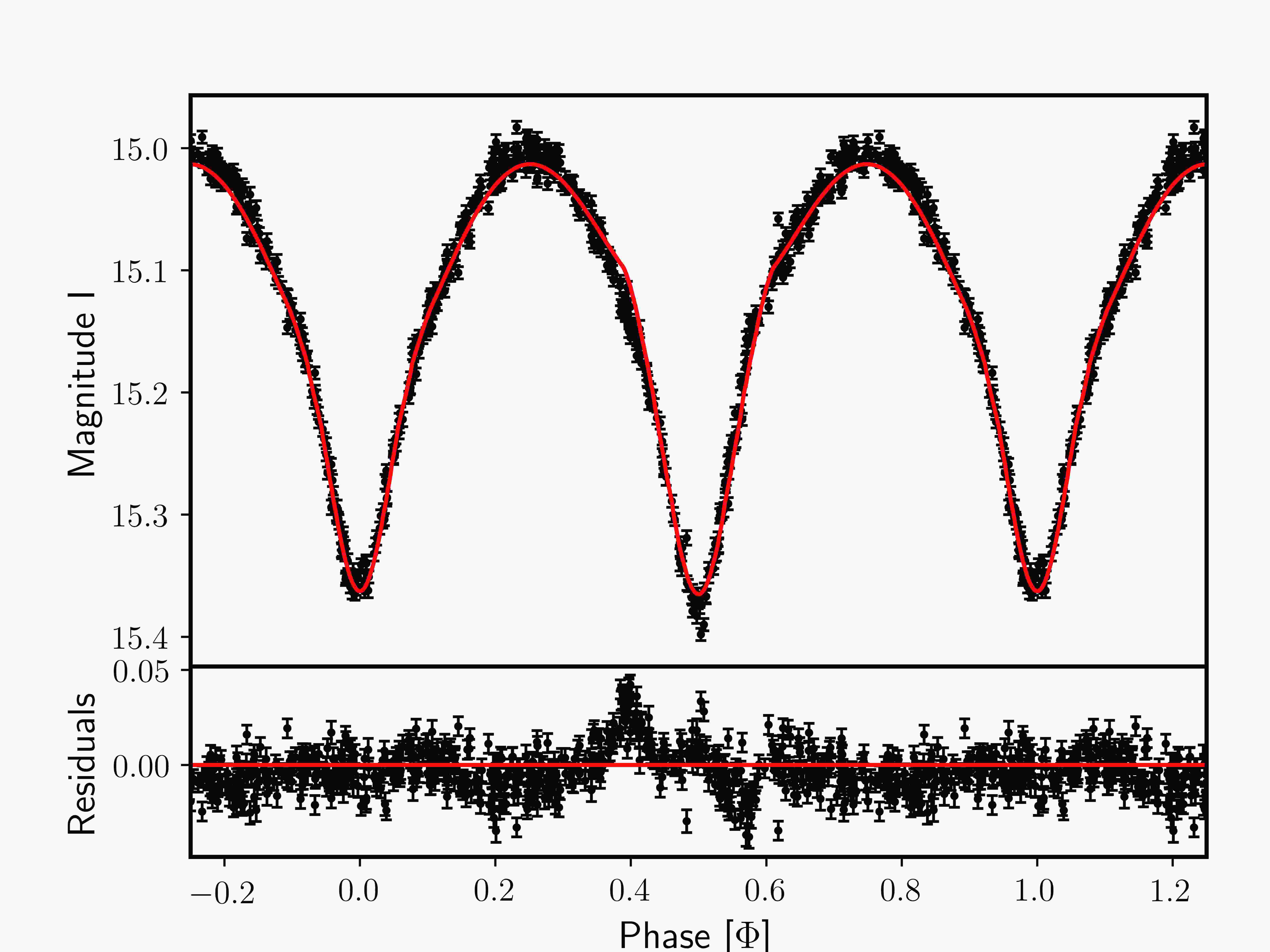}
    \caption{Left: $V$-band light curve of VFTS\,061, phased to the best-fit model derived with PHOEBE. Right: Same for the $I$-band light curve. Residuals of the fits are shown in the lower panels.}\label{fig:061} 
\end{figure*}

VFTS\,061 is a semi-detached binary system with a circular orbit and a period of 2.33 days. The two stars were classified as O9\,V and O9\,III in Paper III. We fix the effective temperatures to 33500\,K for the primary and to 32900\,K for the secondary. From PHOEBE we derive an inclination of $69.1\degr \pm 0.9$\degr. The two components have roughly the same temperature even though the secondary has a lower surface gravity, filling in its Roche lobe. We estimate a mass of $M_P = 16.3 \pm  1.4\,\msun$ and a mean radius of $R_P = 7.2 \pm 0.2\,\rsun$ for the primary whilst the secondary has a mass of $M_S = 8.7 \pm 0.6\,\msun$ and a radius of $R_S = 7.3 \pm 0.3\,\rsun$. Figure\,\ref{fig:061} displays the light curve and its PHOEBE best-fit model.

\subsubsection{VFTS\,094}
\label{result:VFTS094}

VFTS\,094 is an O4$+$O6 binary system with very low projected masses for both components ($M_P \sin^3 i = 3.19 \pm 0.59\,\msun$ and $M_S \sin^3 i = 2.95 \pm 0.54\,\msun$) and an orbital period of 2.26 days. We note that no luminosity class was assigned to stars with spectral classification earlier than O7 in Paper III because of the wavelength range covered by the spectroscopic observations. However, \citet{walborn14} classified the primary as O3.5\,Inf$^{*}$p. Its light curve (Fig.\,\ref{fig:094}) displays ellipsoidal variations with $\Delta m = 0.05 $ in the $V$ band and $\Delta m = 0.08$ in the $I$ band. In Paper III we derived effective temperatures of 41900\,K for the primary and 40100\,K for the secondary. From PHOEBE we estimate for the system an inclination of $i=28.1_{-3.0}^{+3.4}$\degr. This yields masses of $M_P=30.5_{-10.6}^{+11.6}$\,\msun\ and $M_S = 28.2_{-9.8}^{+10.7}$\,\msun\ as well as radii of $R_P=10.9_{- 1.7}^{+ 1.1}\,\rsun$ and $R_S=8.1_{- 1.1}^{+ 1.2}\,\rsun$.

\subsubsection{VFTS\,176}
\label{result:VFTS176}

VFTS\,176 is classified as an O6 $+$ B0.2\,V binary system with an orbital period of 1.78 days. The effective temperature of the primary is fixed to 38300\,K and that of the secondary to 28500\,K. Its light curves show eclipses (Fig.\,\ref{fig:176}). We determine an inclination of $i = 89.70\degr \pm 0.30\degr$. We estimate that the primary has a radius of $ R_P = 9.4 \pm 0.1\,\rsun$, filling in its Roche lobe.  The secondary has a radius of $R_S = 4.9 \pm 0.2\,\rsun$. We estimate masses of $M_P = 28.3 \pm 1.5\,\msun$ and $M_S = 17.5 \pm 0.9\,\msun$. 

 \subsubsection{VFTS\,450}
\label{result:VFTS450}

VFTS\,450 was analysed by \citet{howarth15}. The fainter component (the primary) is classified as O8\,V and the brighter one (the secondary) as O9.7\,I. The system has an orbital period of 6.89 days. We fix the effective temperatures to 33800\,K for the primary and 28300\,K for the secondary. The light curve of VFTS\,450 (Fig.\,\ref{fig:450}) shows ellipsoidal variations and perhaps grazing eclipses, but the dispersion of the data points prevents us from being more precise. Given the spectral classification of the two components, the most plausible configuration is that the secondary fills its Roche lobe. Under this assumption, we determine an inclination of about $63.5_{-1.2}^{+1.7}$\degr. This inclination infers masses of $M_P = 29.0_{- 4.0}^{+ 4.1}\,\msun$ for the primary and of $M_S = 27.8_{- 3.8}^{+ 3.9}\,\msun$ for the secondary. The radius of the primary is $R_P = 13.0_{-0.6}^{+3.0}\,\rsun$ and that of the secondary $R_S = 22.2_{-0.4}^{+0.3}\,\rsun$. In the situation where both components fill their Roche lobe (contact configuration), the inclination of the system is estimated to be $58.4\degr \pm 1.5\degr$, giving masses of $M_P = 38.2 \pm 0.8\,\msun$ and $M_S = 35.1 \pm 0.6\,\msun$ as well as radii of $R_P = 26.6 \pm 0.5\,\rsun$ and $R_S = 25.7 \pm 0.4\,\rsun$. 

\subsubsection{VFTS\,652}
\label{result:VFTS652}

VFTS\,652 is a binary system composed of an O8\,V primary and an evolved B1\,I secondary, and was already studied by \citet{howarth15}. The orbital period of the system is 8.59 days. The configuration of the system indicates that the secondary fills its Roche lobe. However, the light curve only displays ellipsoidal variations and perhaps grazing eclipses (Fig.\,\ref{fig:652}), but the dispersion of the data points does not allow us to distinguish between the two possibilities. PHOEBE suggests an inclination of $i=63.7_{-4.8}^{+0.9}$\degr. From this value and from effective temperatures of 32100\,K for the primary and 23900\,K for the secondary, we determine masses of $M_P = 18.1_{- 4.2}^{+ 3.6}\,\msun$ and of $M_S = 6.5_{- 1.1}^{+ 0.8}\,\msun$. We also estimate radii of $R_P = 15.4_{- 0.3}^{+ 0.7}\,\rsun$ and $R_S = 16.8_{-0.2}^{+0.7}\,\rsun$. We note that these results are in agreement with those of \citet{howarth15} within the error bars. 

\subsection{Contact systems}

\subsubsection{VFTS\,066}
\label{result:VFTS066}

VFTS\,066 is a system classified as O9\,V$+$B0.2\,V. It has an orbital period of 1.14 days. Its light curve (Fig.\,\ref{fig:066}) displays ellipsoidal variations due to the deformation of the stars. The minimum masses of $M_P \sin^3 i = 0.35 \pm 0.01\,\msun$ and of $M_S \sin^3 i = 0.18 \pm 0.01\,\msun $ given in Papers I and II suggest a very low inclination for the system. From Paper III, the effective temperature of the primary is fixed to 32800\,K and to 29000\,K for the secondary. With its short orbital period, we can assume that the two objects are close to filling or fill their Roche lobe. The fit of the light curve confirms these assumptions. We indeed determine an inclination of $i = 17.5_{-2.5}^{+3.2}$\degr, which provides masses of $M_P =13.0_{- 5.0}^{+ 7.0}\,\msun$ for the primary and $M_S = 6.6_{- 2.8}^{+ 3.5}\,\msun$ for the secondary. Their radii are estimated to $R_P=5.8_{- 0.8}^{+ 0.5}\,\rsun$ and $R_S = 4.4_{- 0.8}^{+ 0.4}\,\rsun$, respectively. This analysis also confirms that, within the error bars on the inclination, both components fill or exceed their Roche lobes. 

\subsubsection{VFTS\,352}
\label{result:VFTS352}

The light curve of VFTS\,352 (Fig.\,\ref{fig:352}) has been thoroughly analysed in photometry by \citet{almeida15} and in spectroscopy by \citet{abdul-masih19}. We derive values for the effective temperatures of both components of 41600\,K for the O5.5\,V primary and 40600\,K for the O6.5\,V secondary. VFTS\,352 has an orbital period of 1.12 days. The light curve shows an over-contact system seen under an inclination of $53.6_{-0.9}^{+1.3}$\degr. We derive masses of $M_P = 25.6_{- 1.4}^{+ 1.7}\,\msun$ for the primary and of $M_S = 25.1_{- 1.4}^{+ 1.6}\,\msun$ for the secondary. Their mean radii are estimated to $R_P = 6.8_{-0.2}^{+0.1}\,\rsun$ and $R_S = 6.8_{-0.2}^{+0.1}\,\rsun$, respectively. Taking the effective temperature from \citet{abdul-masih19} into account leads to a slightly higher inclination of $55.5\degr \pm 0.7\degr$, masses of $M_P = 23.7_{- 1.3}^{+ 1.6}\,\msun$ and $M_S = 23.3_{- 1.2}^{+ 1.4}\,\msun$ and radii of $R_P = 6.4_{-0.2}^{+0.1}\,\rsun$ and $R_S = 6.3_{-0.2}^{+0.1}\,\rsun$. These results are consistent with our study. However, we obtain smaller radii and lower masses than \citet{almeida15} derived from their analysis, although the differences are small. 

\subsection{Uncertain configurations}

\subsubsection{VFTS\,217}
\label{result:VFTS217}

VFTS\,217 is a binary system composed of an O4 primary and an O5.5 secondary with an orbital period of 1.86 days. Given their spectral types and the short orbital period of the system, we expect that both components almost fill their Roche lobe. The light curve of the system displays ellipsoidal variations (Fig.\,\ref{fig:217}), but no eclipses. The fit of the light curve provides an inclination of the system of $40.0 \pm 4.0$\degr. With effective temperatures of 45000\,K and 41800\,K for the primary and the secondary, respectively, we estimate that the mass and radius of the primary are $M_P= 46.8 \pm 11.7\,\msun$ and $R_P=10.1_{- 1.2}^{+ 1.5}\,\rsun$, and for the secondary $M_S= 38.9 \pm 9.7\,\msun$ and $R_S=9.4_{- 1.0}^{+ 1.4}\,\rsun$. More than 80\% of the Roche lobe volumes are filled by the stars.

\subsubsection{VFTS\,563}
\label{result:VFTS563}

VFTS\,563, composed of two O9.5\,V stars, shows ellipsoidal variations in its light curve (Fig.\,\ref{fig:563}). The effective temperature is 32400\,K for both stars. PHOEBE provides us with an inclination for the system of $i = 29.7_{-2.0}^{+4.9}$\degr. The masses that we infer from this inclination are $M_P = 26.2_{-5.2}^{+ 11.9}\,\msun$ for the primary and $M_S = 20.0_{- 3.9}^{+ 9.1}\,\msun$ for the secondary. Their radii are computed to be $R_P = 6.6_{-0.7}^{+0.4}\,\rsun$ and $R_S = 5.8_{-0.6}^{+0.4}\,\rsun$, meaning that about 85\% of the Roche lobe volumes are filled by the stars.

\section{Discussion}
\label{discussion}

\subsection{Mass-luminosity relation}
\label{relation}

\begin{figure*}
  \centering
    \includegraphics[trim=0 5 0 0,clip,width=9cm]{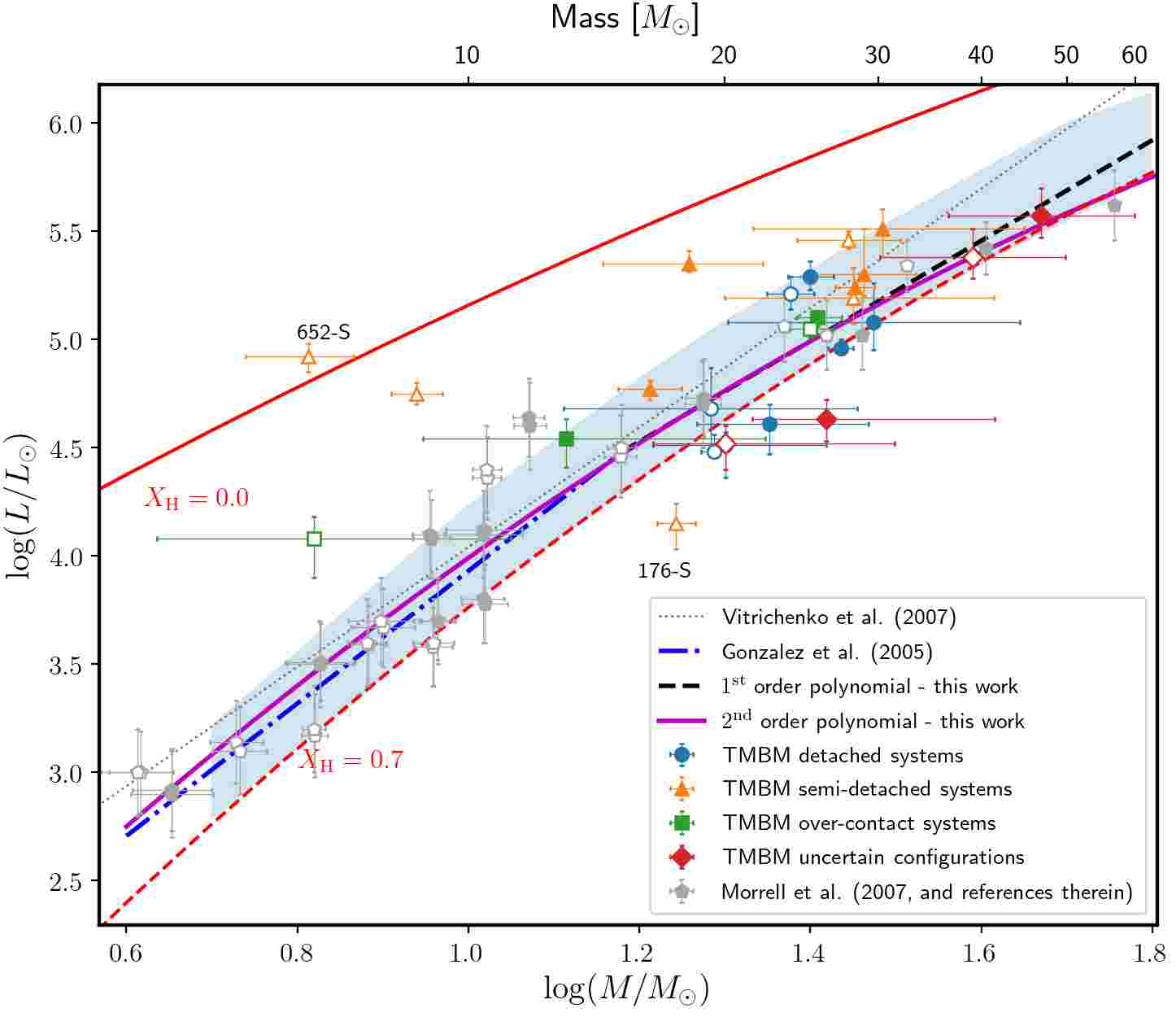}
    \includegraphics[trim=0 5 0 0,clip,width=9cm]{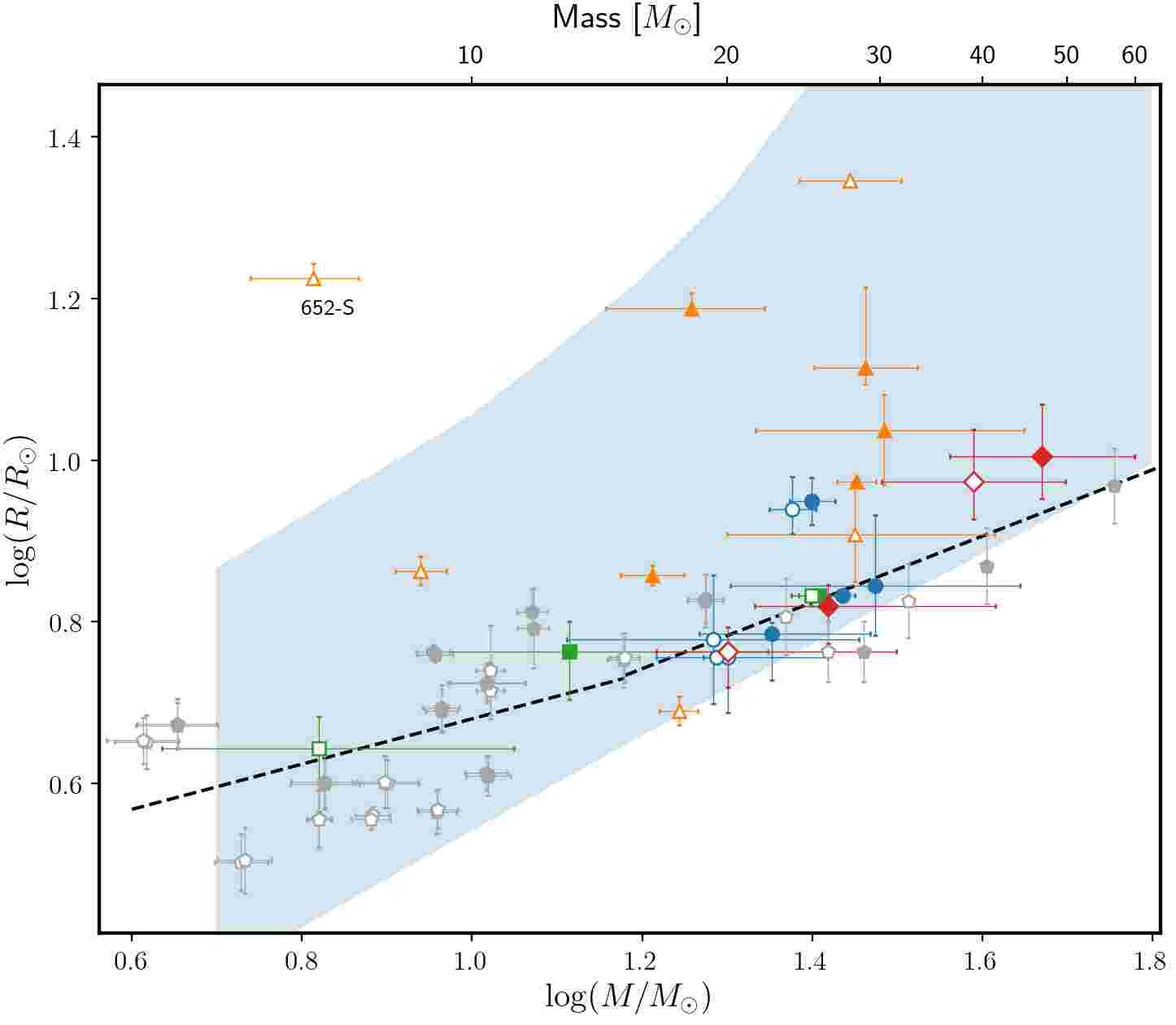}
    \caption{{\it Left: } Mass-luminosity diagram of all the components in our sample and from \citet[][and references therein]{morrell07}. Detached systems are represented in blue, semi-detached systems in orange, over-contact systems in green, and systems with uncertain configurations in red. The grey dots represent the sample of Morrell et al. The blue shaded zone indicates the main sequence from the tracks of \citet{brott11} and \citet{kohler15}, computed with initial rotational velocity equal to $150$\,\kms. Overplotted in dashed and solid lines are the mass-luminosity relations of \citet{graefener11} for chemically homogeneous stars with a hydrogen mass fraction of $X_H = 0.7$ and $X_H = 0$. {\it Right:} Mass-radius diagram built from our sample and from the sample of \citet{morrell07}. The colour-coding is the same as in the left panel.}\label{fig:mass-lum} 
\end{figure*}

The analysis of the 13 light curves of O- and early B-type binary systems in 30\,Dor allows us to determine the masses and the radii of their components. Even though not all our systems are clearly detached, we are able to establish the mass-luminosity and mass-radius relations for massive stars in 30\,Dor, and we compare them with other relations found for the LMC, SMC, or the Milky Way. There is, however, a considerable spread expected between the different mass ranges, even for `well-behaved' single stars. According to the  theory, the mass-luminosity relation follows $$ L \sim M^{\alpha}. \mu^{\beta}, $$  where $\mu$ is the mean molecular weight of the stellar gas and $\beta > 1$ (for a star with 30\,\msun, $\beta = 2.8$; see Fig.\,17 of \citealt{kohler15}). In this power law the exponent $\alpha$ describes the slope of the mass-luminosity relation in the $\log L - \log M$ plane. This relation is very steep near the solar mass ($\alpha \sim 5$); instead,  $\alpha \rightarrow 1$ when $M \rightarrow \infty$, due to the increasing radiation pressure for higher masses \citep{kippenhahn90}. In the core of the star, $\mu$ can change by a factor of 2.2 during the hydrogen burning phase. Therefore, if the convective core comprises about half of the mass of the star, the average $\mu$ can change by a factor of 1.6, and thus  a spread in luminosity at 30\,\msun\ of $1.6^{\beta}$ is expected, which gives a factor of 3.7 (0.57 dex in logarithm of the luminosity,  \citealt{brott11,kohler15}).

To build these relations for the stars in the LMC, the conservative approach is to first exclude all the systems that are believed to have interactions between the two components (semi-detached or contact systems). From Paper III there is no indication that the remaining systems (VFTS\,217, VFTS\,500, VFTS\,543, VFTS\,563, VFTS\,642, and VFTS\,661) have indeed interacting components. In restricting the sample to these systems, we limit ourselves to a mass range between 20 and 50\,\msun. We then fit to the data a linear regression which gives 
\begin{equation}
\log (L/\lsun) = [2.32 \pm 0.18] \log (M/\msun) + [1.74 \pm 0.08]
\end{equation}
for masses between 20 and 50\,\msun\ (Fig.\,\ref{fig:mass-lum}, right part of dashed black line). When we overplot the individual properties of components in eclipsing binary systems from the LMC analysed by \citet[][and references therein]{morrell07}, we see that their most massive systems agree well with those of our sample and that the relation provided for the mass-luminosity also fits their data. If we now perform a linear regression on their systems with masses between 4 and 20\,\msun, the relation that we obtain is: 
\begin{equation}
\log (L/\lsun) = [3.06 \pm 0.22] \log (M/\msun) + [0.87 \pm 0.11]
\end{equation} 
(confirming that provided by \citealt{gonzalez05}; see Fig.\,\ref{fig:mass-lum}, blue dot-dashed line). Thus, the two relations seem different and clearly depend on the mass range. 

Another mass-luminosity relation was given by \citet{north10} for objects with masses between 4 and 20\,\msun\ located in the SMC. These authors provide a mass-luminosity relation:
\begin{equation}
\log (L/\lsun) = [3.04 \pm 0.11] \log (M/\msun) + [0.90 \pm 0.09].
\end{equation} 
This range is similar to that given by \citet{gonzalez05} and \citet{morrell07} for stars in the LMC, and we conclude that the two relations agree well with each other. 

The mass-luminosity relation was also derived for stars with different mass ranges in the Milky Way. From a compilation of results, \citet{vitrichenko07} derived a mass-luminosity relation of 
\begin{equation}
\log (L/\lsun) = [2.76 \pm 0.02] \log (M/\msun) + [1.28 \pm 0.02]
\end{equation}
for Galactic stars between 10 and 50\,\msun. We display this linear fit in Fig.\,\ref{fig:mass-lum} (dotted line) to compare it with our data. This relation seems to be more an average value than a good fit of the observed data, and emphasises the need for either splitting the mass range to properly fit the data or to increase the degree of the polynomial fit. Another possible difference arises because these authors have not discarded those systems showing probable mass transfer between both components. 

Given our results and those from the literature, it seems that metallicity does not play a role in the mass-luminosity relation, which is not what we expected. A star with the same mass but with a lower metallicity would indeed be expected to have a higher luminosity due to lower opacities and due to the radiation that would be freely flowing to the stellar surface. Moreover, we can confirm from observation that the mass-luminosity relation is different depending on the mass range in question. It indeed shows that the exponent tends to be smaller when the stellar masses are higher. In order to derive a unique relation for the entire mass range of massive stars (from 4 to 50\,\msun) in the LMC, we use a polynomial of degree two. We obtain the following more general relation: 
\begin{equation}
\begin{multlined}
 \log (L/\lsun) = [-0.68 \pm 0.11] (\log (M/\msun))^2 + \\ [4.11 \pm 0.15] (\log (M/\msun)) + [0.55 \pm 0.08 ].
 \end{multlined}
\end{equation}  
This relation is represented as a purple solid line in Fig.\,\ref{fig:mass-lum}. In Fig.\,\ref{fig:mass-lum}, we also plot in red two mass-luminosity relations for homogeneous ZAMS ($X_H = 0.7$) and TAMS ($X_H = 0.0$) stars provided by \citet{graefener11}. Based on the mass-luminosity relation, there can be no core hydrogen burning star which is not located between the two lines (i.e. this  should also be a hard limit  for post-interaction binaries). We see that among the semi-detached binaries the two secondary components of VFTS\,176 and VFTS\,652 are outside these two lines. We also see that two other stars deviate more than $1\sigma$ from these two lines: the primary component of VFTS\,563 and the secondary star of VFTS\,661.

\subsection{Mass discrepancy}
\label{discussion:mass}

Historically, three different approaches\footnote{Actually, a fourth value, the wind mass, was mentioned by \citet{groenewegen89} and \citet{kudritzki92}. This method employs the wind-driven theory, which relates the terminal wind velocity to the stellar escape velocity. This parameter is unknown for the stars in our sample, and so we do not discuss it in the present paper. More information can be found in \citet{weidner10}. } to measuring the mass of a star have been developed:
\begin{itemize}
\item  the spectroscopic mass ($M_{\rm spec}$) is computed from the surface gravity and the mean stellar radius, through atmosphere modelling and absolute luminosity;
\item the evolutionary mass ($M_{\rm evol}$) is obtained from the position of the star in the Hertzsprung-Russell diagram;
\item the dynamical mass ($M_{\rm dyn}$) is  determined from the minimum mass of the component in a binary system and the inclination of a system (estimated through photometry or relative astrometry).
\end{itemize} 

For our analysis, the evolutionary masses were computed by using the Bayesian tool BONNSAI \citep{schneider14,schneider17} with the luminosities, effective temperatures, surface gravities, and projected rotational velocities (derived in Paper III) as inputs. This allows  a statistical comparison of the empirical results with evolutionary tracks of \citet{brott11} and \citet{kohler15}. \citet{herrero92} noted a systematic discrepancy between the evolutionary and the spectroscopic masses with the evolutionary masses on average higher than the spectroscopic estimates. The general picture tends to show now that the evolutionary and spectroscopic masses for populations of massive single stars are roughly consistent for masses higher than 25--35\,\msun\ \citep[see e.g.][]{martins12, mahy15}. In contrast, for objects with initial masses lower than this threshold, the evolutionary masses are systematically higher than the spectroscopic masses \citep[see e.g.][]{markova15, markova18, schneider18b}. 

The analysis of eclipsing detached binary systems provides us with a unique way to determine the exact masses of stars. We can use it as a Rosetta Stone to tackle the mass-discrepancy problem. \citet{martins17} pointed out that the evolutionary masses in their sample of six binary systems were also systematically higher by about 15--20\% than the spectroscopic and dynamical masses. In Table\,\ref{tab:parameters}, we give the dynamical, spectroscopic, and evolutionary masses for each component of the 13 systems with light curves studied in the present paper. We also give the different ratios of these different masses.

\begin{figure}[htbp]
  \centering
    \includegraphics[trim=60 0 75 35,clip,width=8cm]{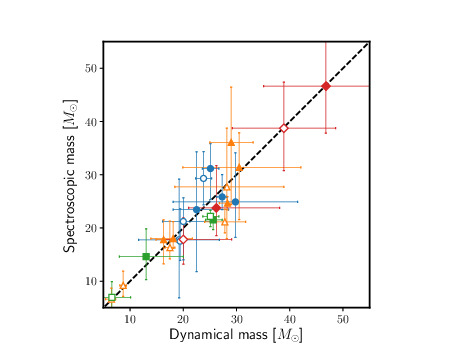}
    \includegraphics[trim=60 0 75 35,clip,width=8cm]{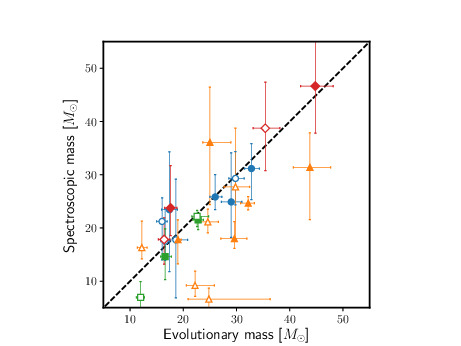}
    \includegraphics[trim=60 0 75 35,clip,width=8cm]{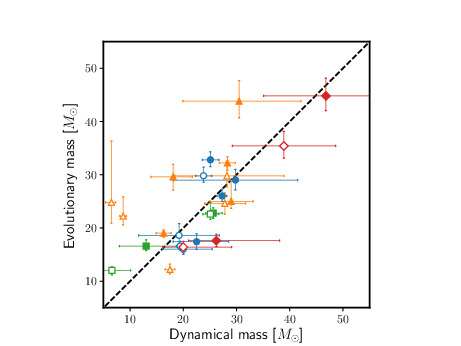}
    \caption{Top: Spectroscopic mass as a function of dynamical mass for the systems observed in photometry. Middle: Spectroscopic mass as a function of evolutionary mass for all the stars of our sample. Bottom: Evolutionary mass as a function of dynamical mass for the systems observed in photometry. The colour-coding is the same as in Fig.\,\ref{fig:mass-lum}.}\label{fig:mass} 
\end{figure}

While dynamical and spectroscopic masses are measurements that are independent of the evolutionary history of binaries, the evolutionary masses depend on the stellar evolution models used to extract them, in particular single-star tracks. A comparison between these three masses is performed in Fig.\,\ref{fig:mass}.

The comparison between the spectroscopic and dynamical masses offers a good agreement within the error bars, even for the components in interacting systems (given that these measurements are independent of the evolution of the stars). Four stars, however, deviate by about $1.5-\sigma$ (see Table\,\ref{tab:parameters}): the primary of VFTS\,176, the two components of VFTS\,352 and the secondary of VFTS\,450. These four objects are either in semi-detached or contact configurations, but as mentioned above their evolution should not have an impact on these two masses. The causes of this discrepancy therefore remain  uncertain.

When we consider the evolutionary versus spectroscopic masses (Fig.\,\ref{fig:mass}, middle panel), we observe that a general agreement exists between these two masses for stars in detached systems. For objects in semi-detached systems, and to a lesser extent in contact systems, the discrepancy is clearly visible, with evolutionary masses systematically higher (i.e. these stars tend to be overluminous). In almost all the case (except for VFTS\,652 where both components are overluminous when we compare the spectroscopic and evolutionary masses), the components that show the largest discrepancy are the stars that fill the Roche lobe, suggesting that the mass transfer and thus the stripping of their outer layers makes them more luminous than their physical properties tend to indicate. The comparison of their properties with single-star evolutionary tracks is therefore no longer valid since mass transfers affect the properties of these components. 

When we compare the dynamical masses to the evolutionary masses, we draw the same conclusions. There are, however, three clear outliers among the detached systems: both components of VFTS\,500, and the secondary of VFTS\,661. As mentioned above, we also observe a systematic shift indicating that the evolutionary masses tend to be higher than the dynamical masses for the components in interacting systems (semi-detached and contact). The stripping of the outer layers of the mass donors tends to make them more luminous, increasing their evolutionary mass. This result confirms that the use of single-star evolutionary tracks to explain the product of binary interactions do not match. This trend was also observed by \citet{martins17} for massive binary systems located in our Galaxy. 

\section{Conclusion}
\label{conclusion}

We   analysed the OGLE light curves of 13 binary systems in the 30 Doradus region. Among the 13 light curves there are  only three that display deep eclipses; the others are reminiscent of ellipsoidal variations or over-contact systems. The study of the light curves gives us access to the inclinations of the systems, and thus to the dynamical masses and radii of each component. However, for the systems showing ellipsoidal variations, the uncertainties on their inclinations are large, which also implies large uncertainties on their masses and radii. These data complement the optical spectra obtained in the frame of the Tarantula Massive Binary Monitoring project \citep{almeida17}. Disentangled spectra were already modelled  with an atmosphere code, and the individual parameters of each component are provided in Paper III. The effective temperatures of each component were adopted here to model the OGLE light curves.

By fitting the light curves, we confirmed the configurations established in Paper III for four detached, four semi-detached, and two contact systems. We also revised these configurations for three systems. For VFTS\,094, the system is expected to be semi-detached. For VFTS\,217 and VFTS\,563, the configurations are uncertain and it is not clear whether these systems are detached or in contact. These uncertainties come from the low inclinations inferred for those systems and because their light curves do not show any eclipses, but do show ellipsoidal variations.

We used the masses, luminosities, and radii of the stars to compare the mass-luminosity and mass-radius relations. As expected from  theory, the relations are different with respect to the mass ranges of the stars, but appear to be independent of the metallicity regimes. If   covering the entire mass range of massive stars (from 4 to 50\,\msun) is required, we provide a second-degree equation that fits our data  best. 

We   also compared these dynamical, spectroscopic, and evolutionary masses of the star in our sample. While the two first masses are measurements, the others are dependent on the evolution of the stars. We observe a general agreement, within the error bars, between the dynamical and spectroscopic masses. The two masses agree well with each other, which implies a high credibility for the dynamical masses and   for the spectroscopic masses for the entire  mass range under investigation. A discrepancy is clearly present when we compare the dynamical and spectroscopic masses to the evolutionary masses, with the largest differences occurring for members of the semi-detached systems.

Through the present analysis we provide a set of stellar parameters for massive binary systems that can now be compared with binary evolutionary models in order to validate these models and to better understand the different stages of evolution of massive binary systems.

\begin{acknowledgements}
We thank the anonymous referee for his/her constructive remarks about the paper. This work was based on observations obtained with the 1.3-m Warsaw Telescope at the Las Campanas Observatory of the Carnegie Institution of Washington. L.A.A. thanks to Aperfei{\c c}oamento de Pessoal de N{\'i}vel Superior (CAPES) and Funda{\c c}{\~a}o de Amparo {\`a} Pesquisa do Estado de S{\~a}o Paulo (FAPESP – 2011/51680-6, 2012/09716-6, 2013/18245-0) for financial support. H.S. acknowledge support from the FWO-Odysseus programme under project G0F8H6NT. H.S. and T.S. acknowledge support from the European Research Council (ERC) under the European Union’s DLV-772225-MULTIPLES Horizon 2020 research and innovation programme. S.dM. acknowledges funding by the European Union's Horizon 2020 research and innovation programme from the European Research Council (ERC) (Grant agreement No.\ 715063), and by the Netherlands Organisation for Scientific Research (NWO) as part of the Vidi research program BinWaves with project number 639.042.728. A.F.J.M. is grateful to NSERC (Canada) for financial aid.
\end{acknowledgements}

\bibliography{TMBM}

\begin{appendix}
  \section{PHOEBE best fits}
  In this appendix, we compare the PHOEBE best-fit models to the $V$ and $I$ light curves of the systems.

\begin{figure*}[htbp]
  \centering
    \includegraphics[width=9cm, angle=0]{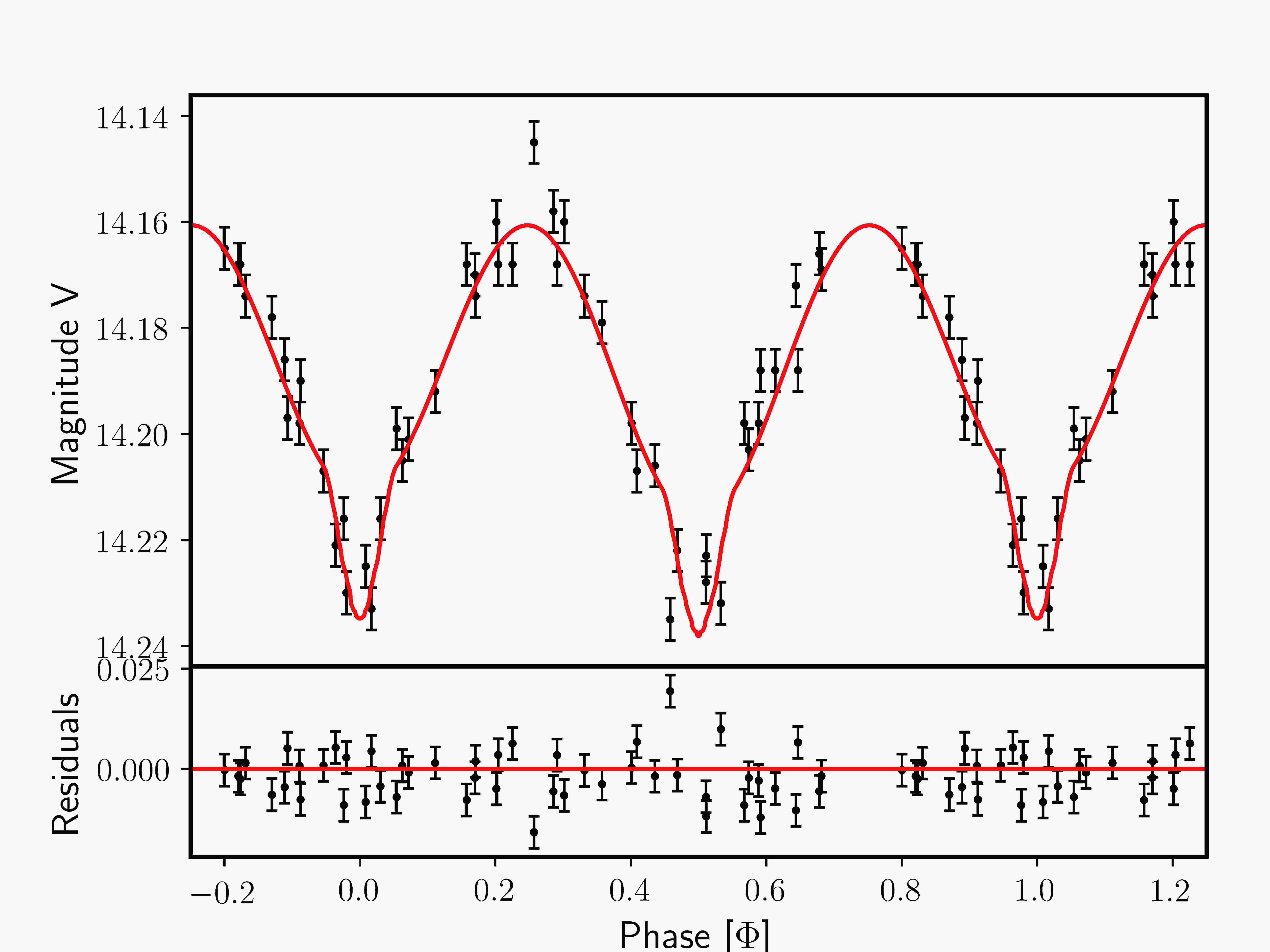}
    \includegraphics[width=9cm, angle=0]{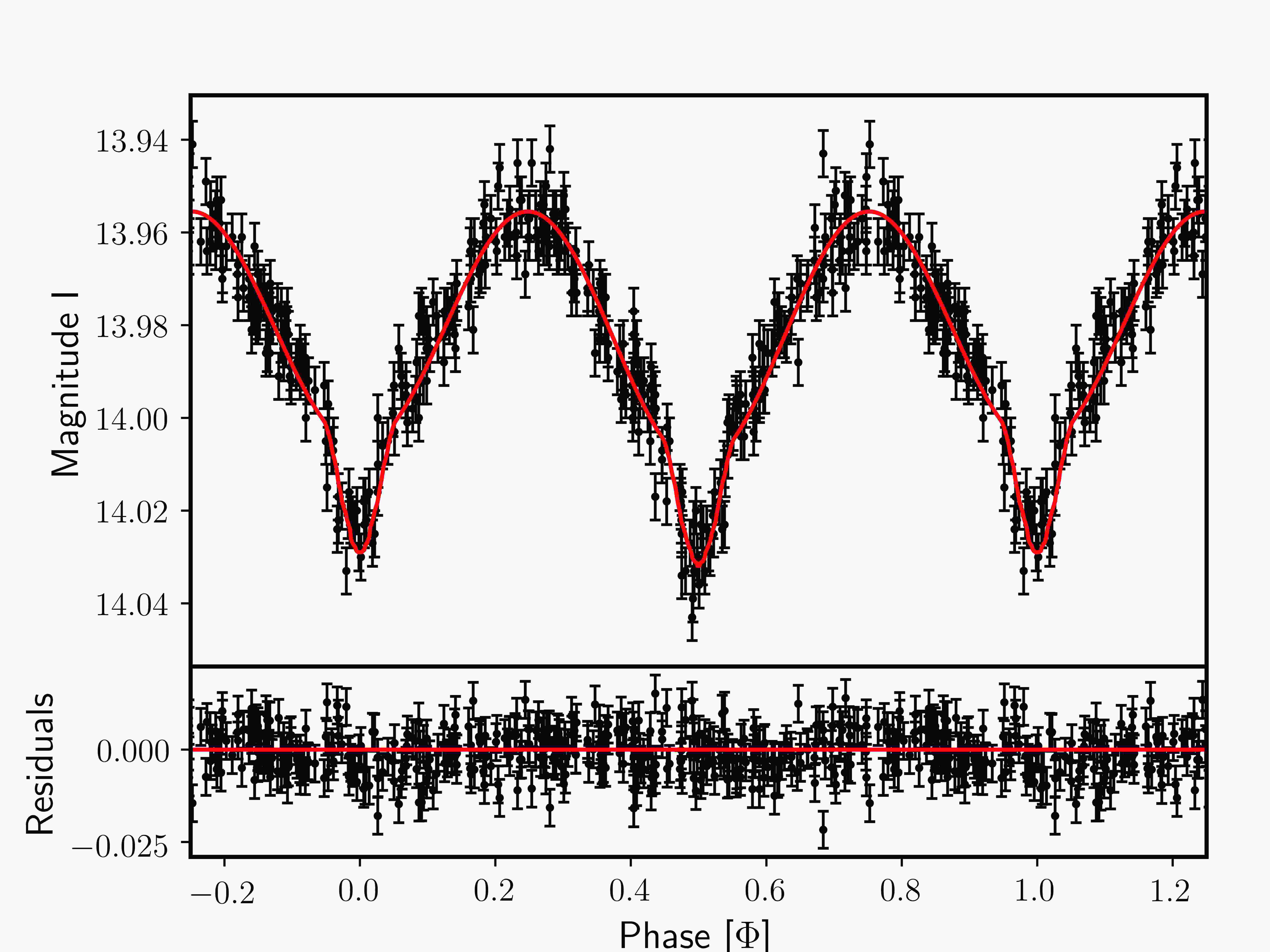}
    \caption{Same as Fig.\,\ref{fig:061}, but for VFTS\,500.}\label{fig:500} 
\end{figure*}

\begin{figure*}[htbp]
  \centering
    \includegraphics[width=9cm, angle=0]{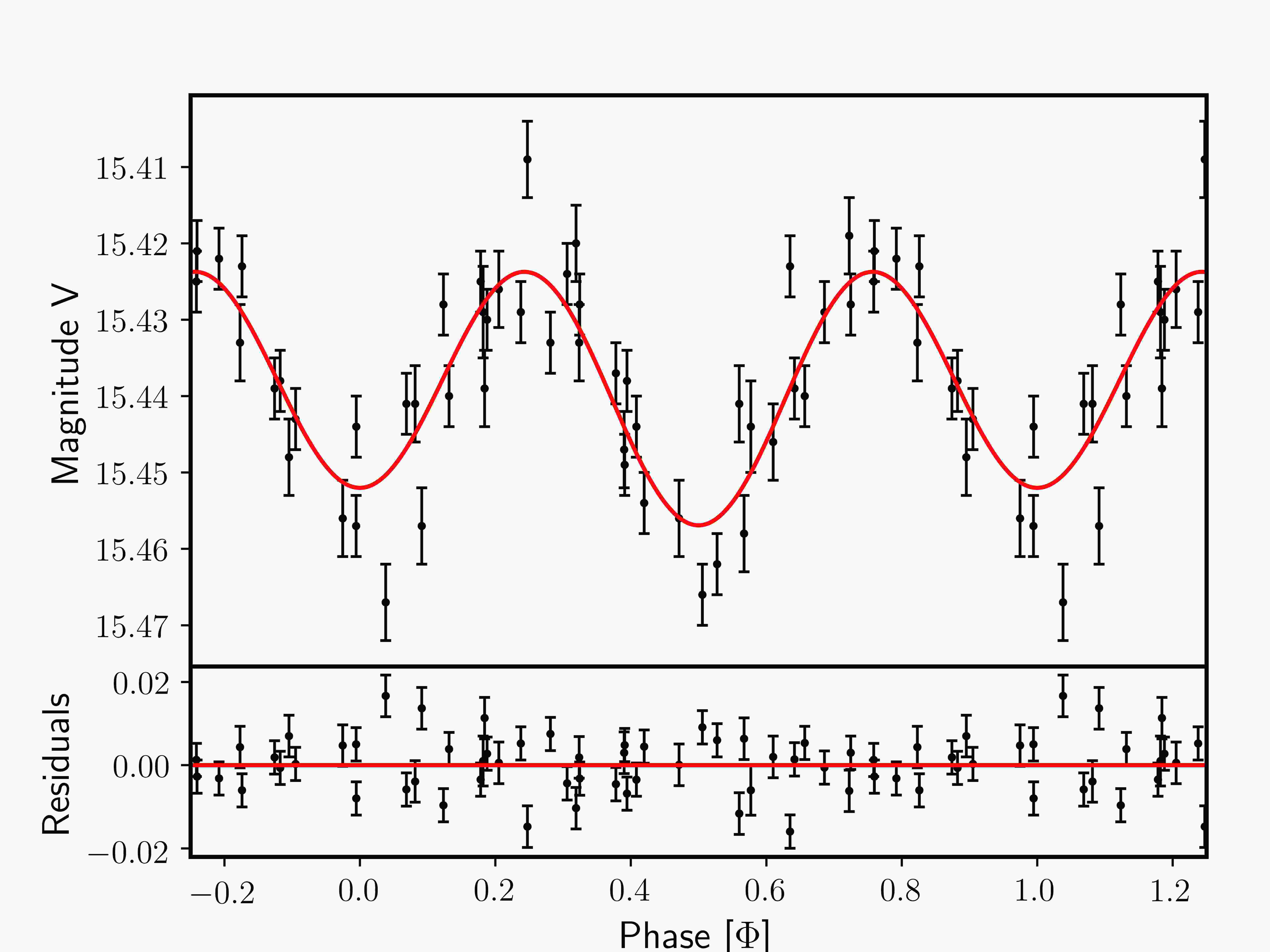}
    \includegraphics[width=9cm, angle=0]{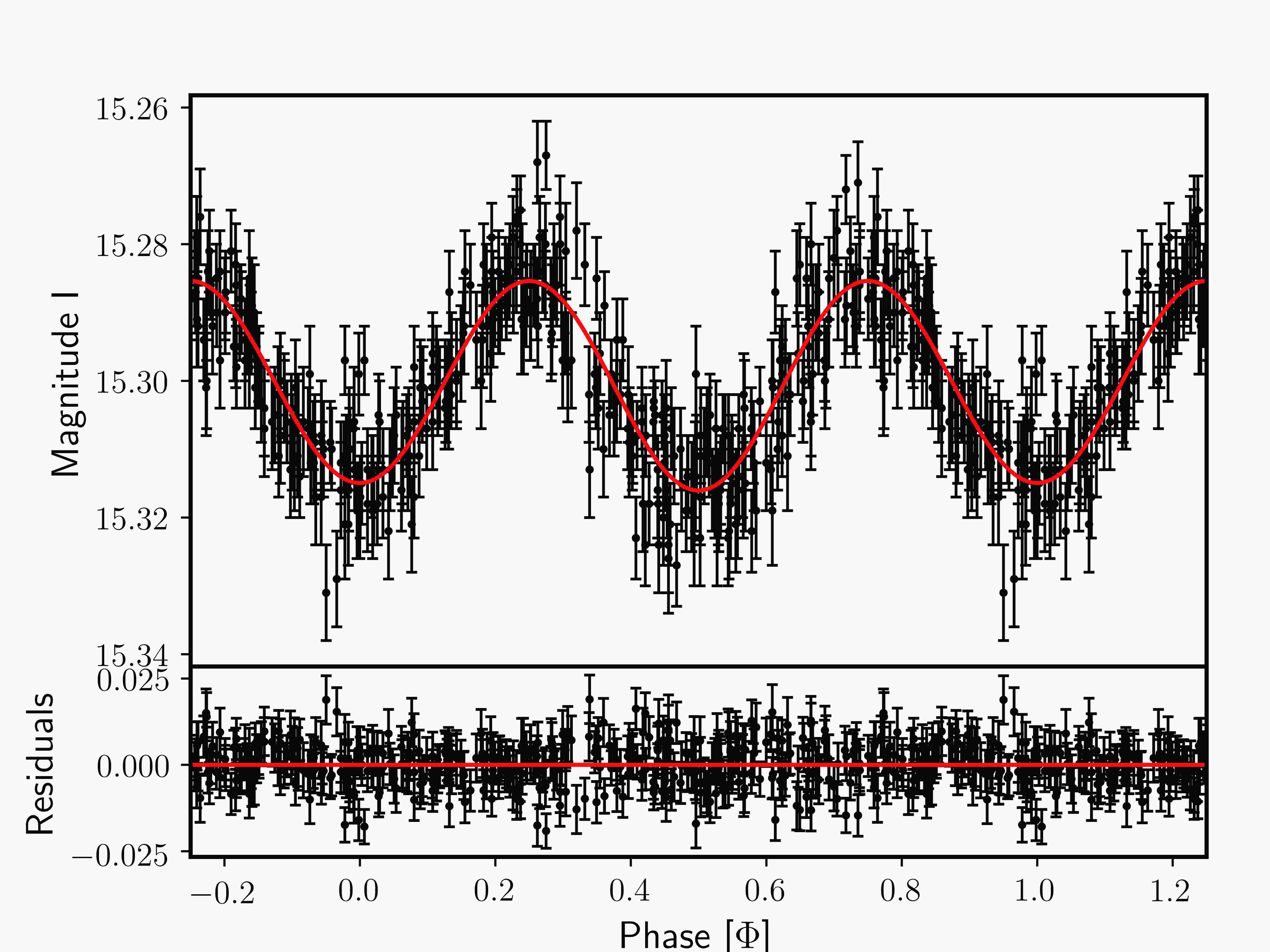}
    \caption{Same as Fig.\,\ref{fig:061}, but for VFTS\,543.}\label{fig:543} 
\end{figure*}

\begin{figure*}[htbp]
  \centering
    \includegraphics[width=9cm, angle=0]{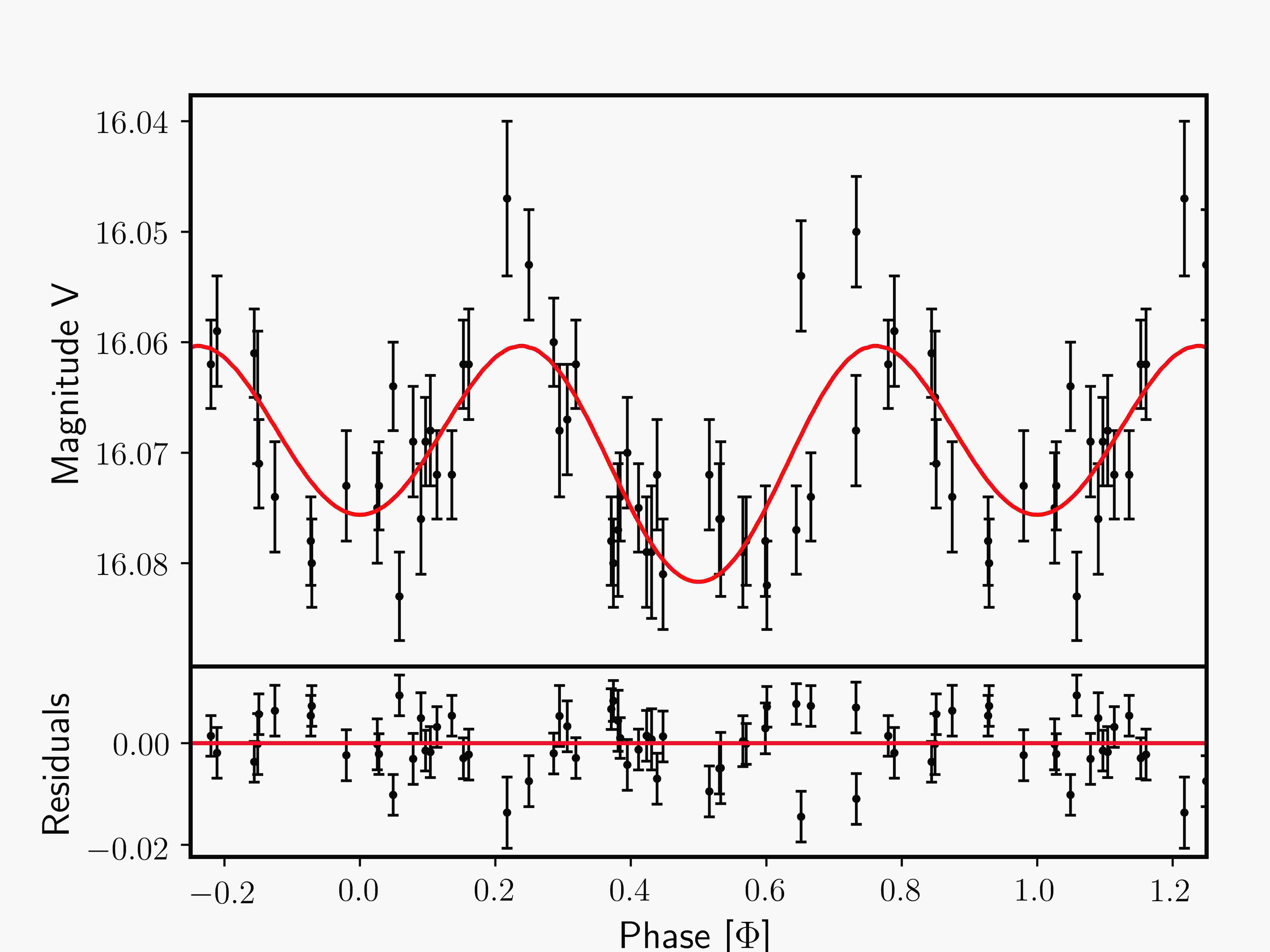}
    \includegraphics[width=9cm, angle=0]{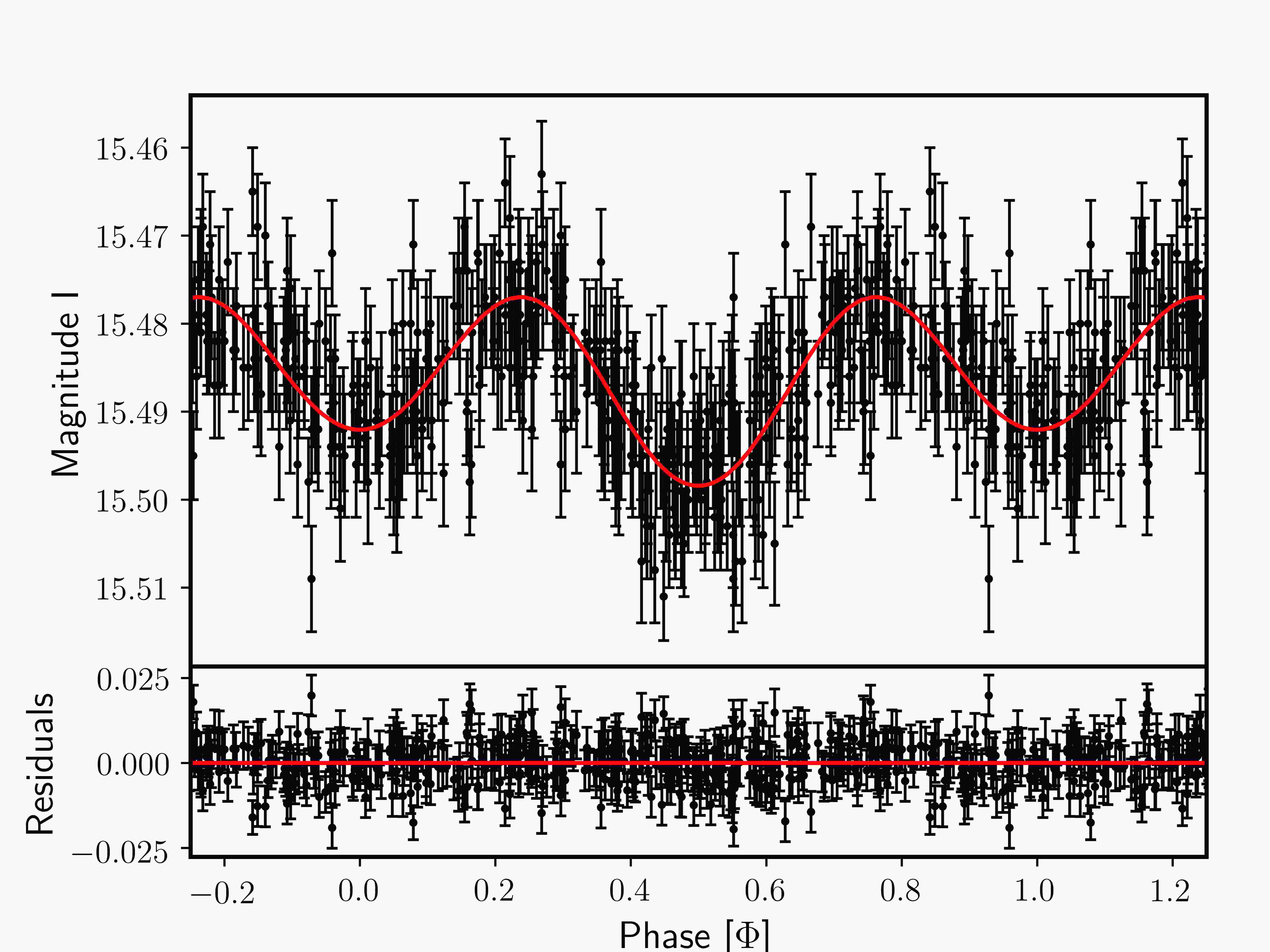}
    \caption{Same as Fig.\,\ref{fig:061}, but for VFTS\,642.}\label{fig:642} 
\end{figure*}

\begin{figure*}[htbp]
  \centering
    \includegraphics[width=9cm, angle=0]{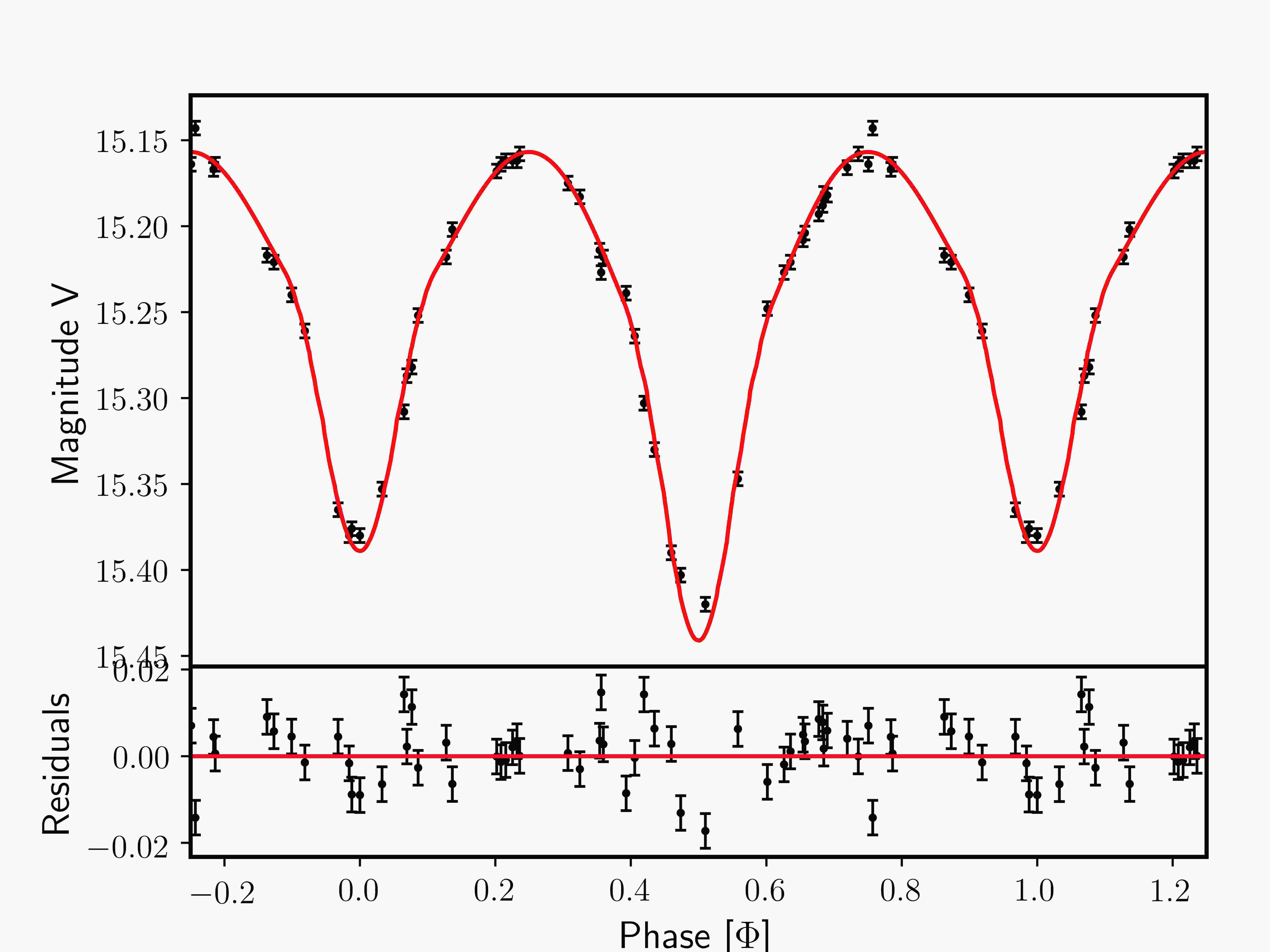}
    \includegraphics[width=9cm, angle=0]{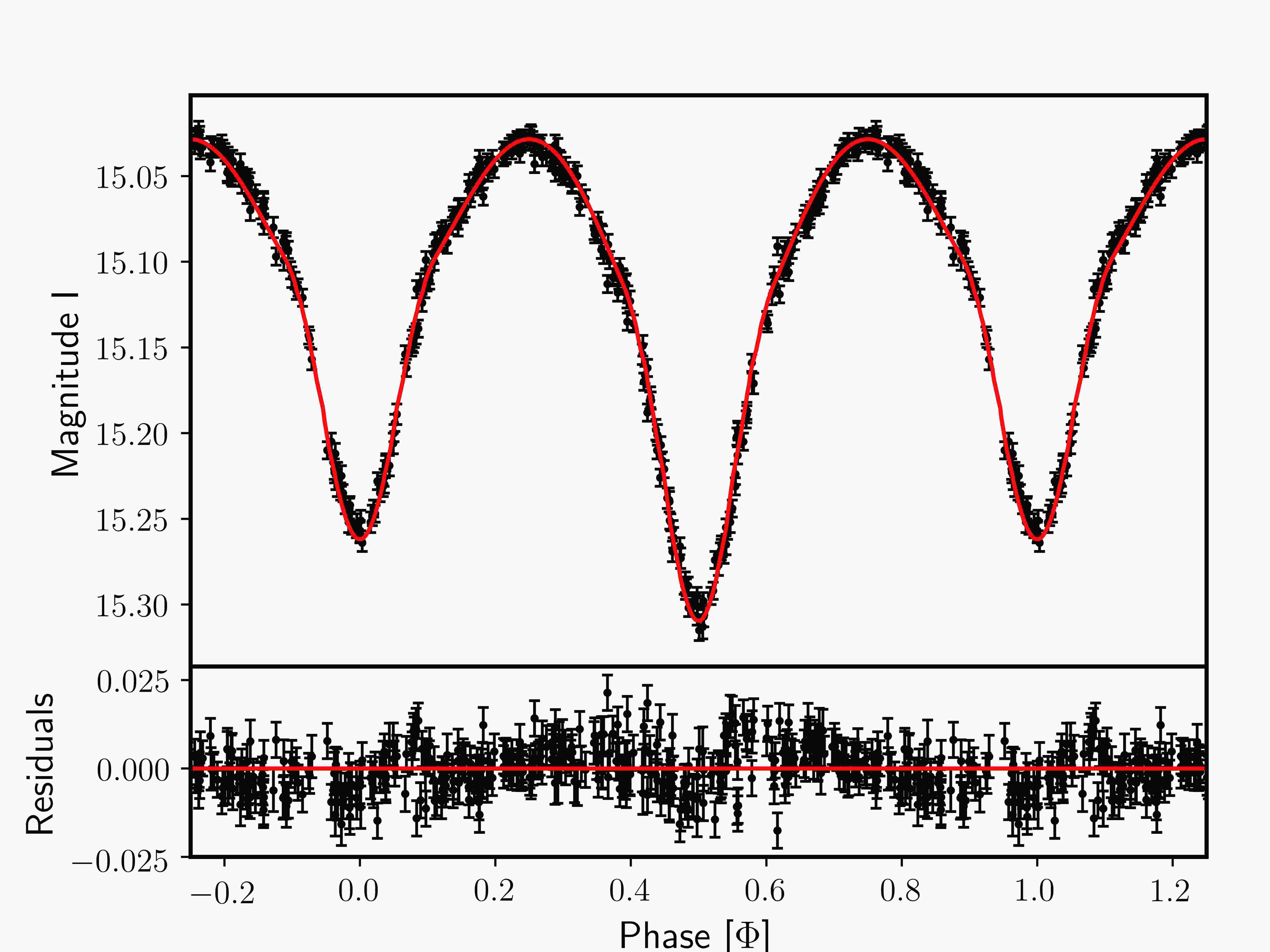}
    \caption{Same as Fig.\,\ref{fig:061}, but for VFTS\,661.}\label{fig:661} 
\end{figure*}

\begin{figure*}[htbp]
  \centering
    \includegraphics[width=9cm, angle=0]{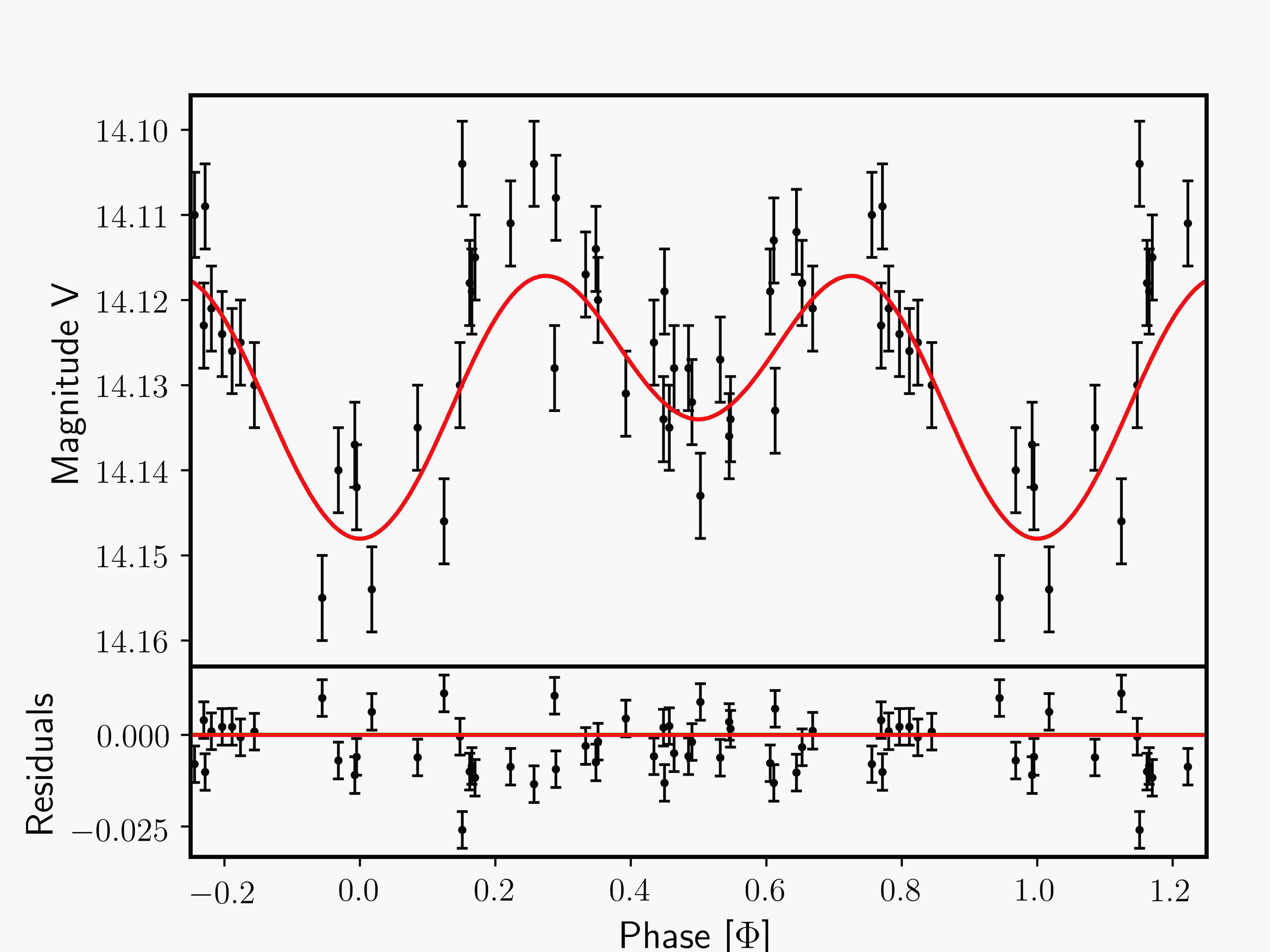}
    \includegraphics[width=9cm, angle=0]{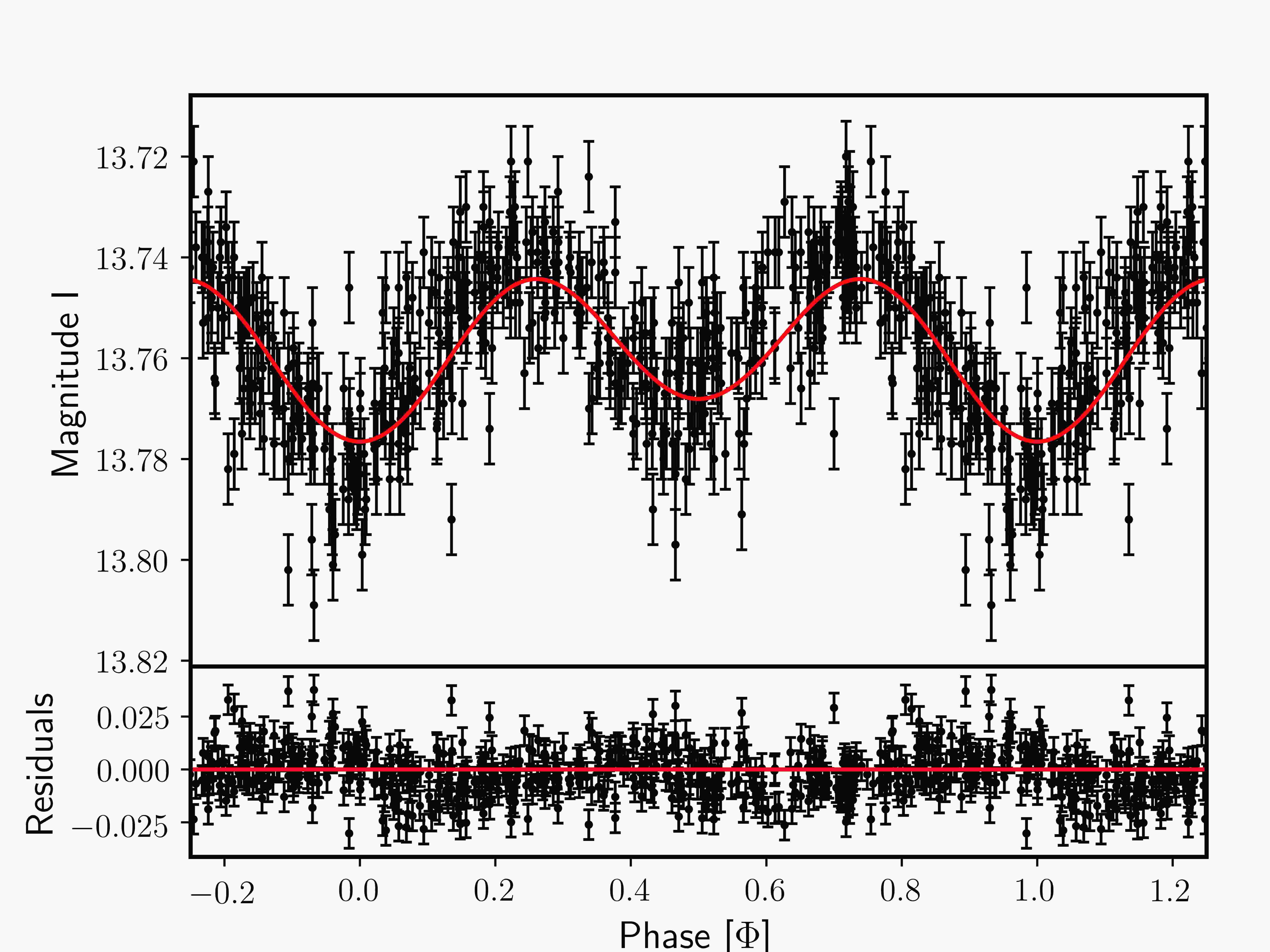}
    \caption{Same as Fig.\,\ref{fig:061}, but for VFTS\,094.}\label{fig:094} 
\end{figure*}  

\begin{figure*}[htbp]
  \centering
    \includegraphics[width=9cm, angle=0]{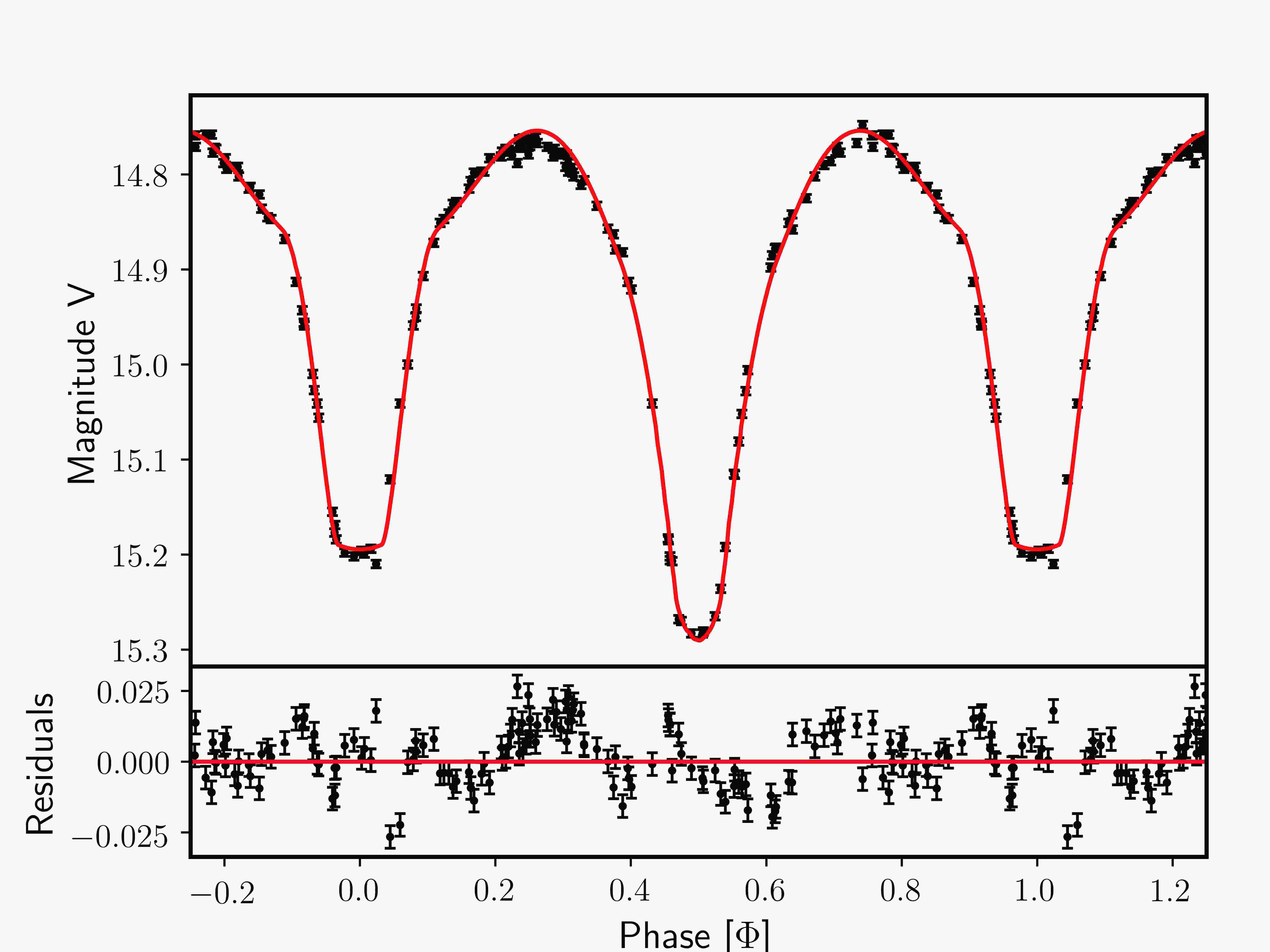}
    \includegraphics[width=9cm, angle=0]{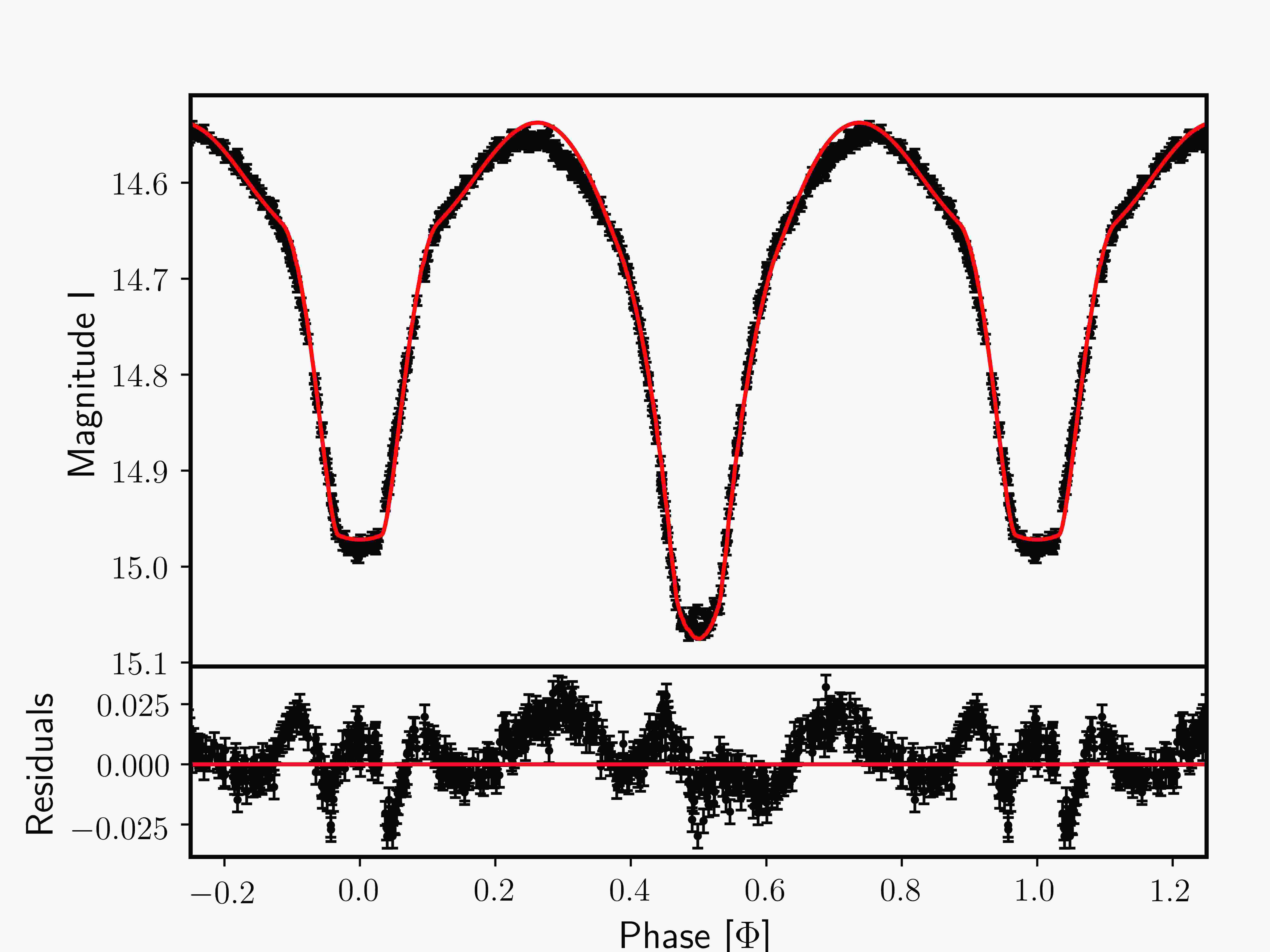}
    \caption{Same as Fig.\,\ref{fig:061}, but for VFTS\,176.}\label{fig:176} 
\end{figure*}

\begin{figure*}[htbp]
  \centering
    \includegraphics[width=9cm, angle=0]{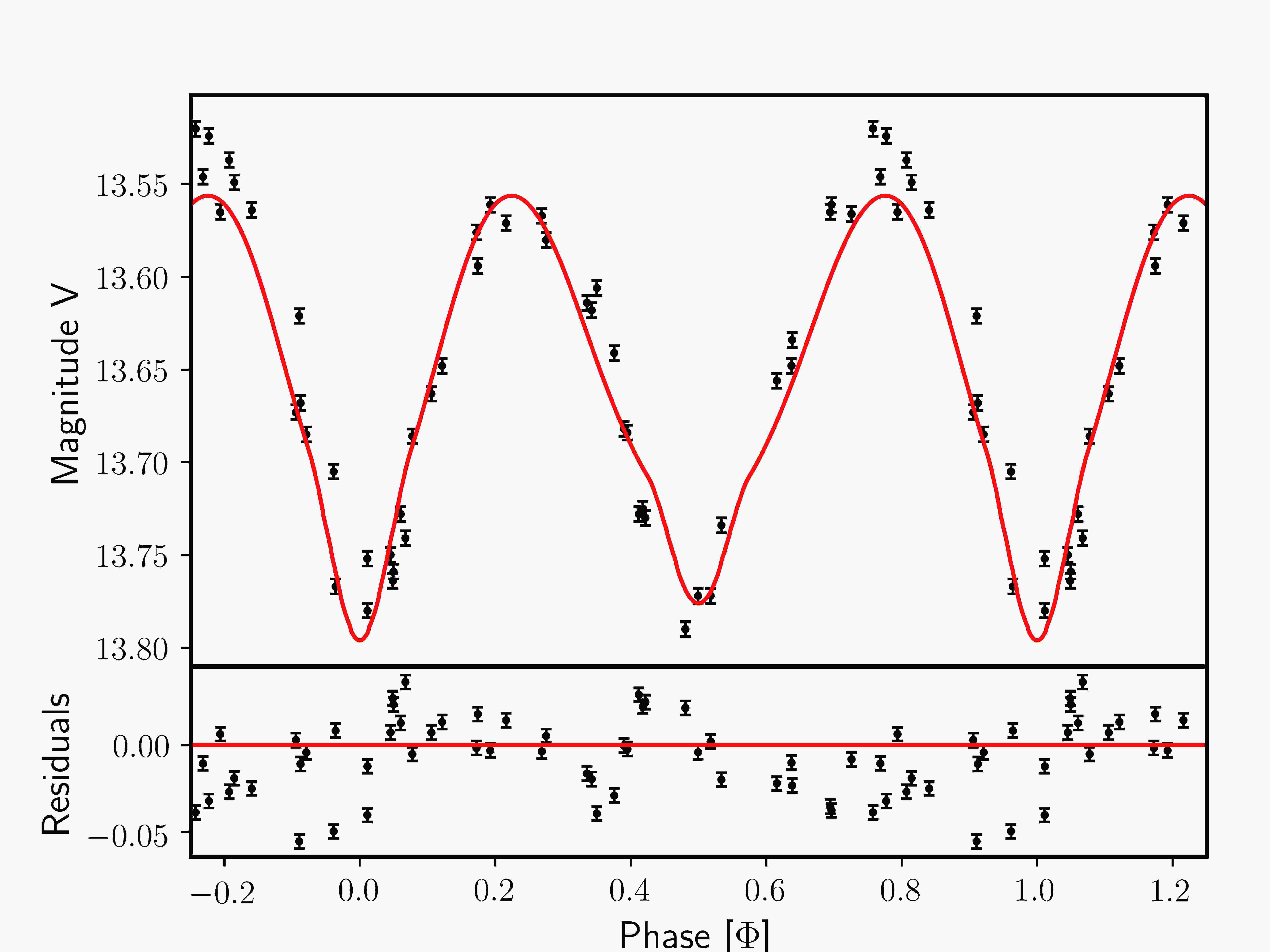}
    \includegraphics[width=9cm, angle=0]{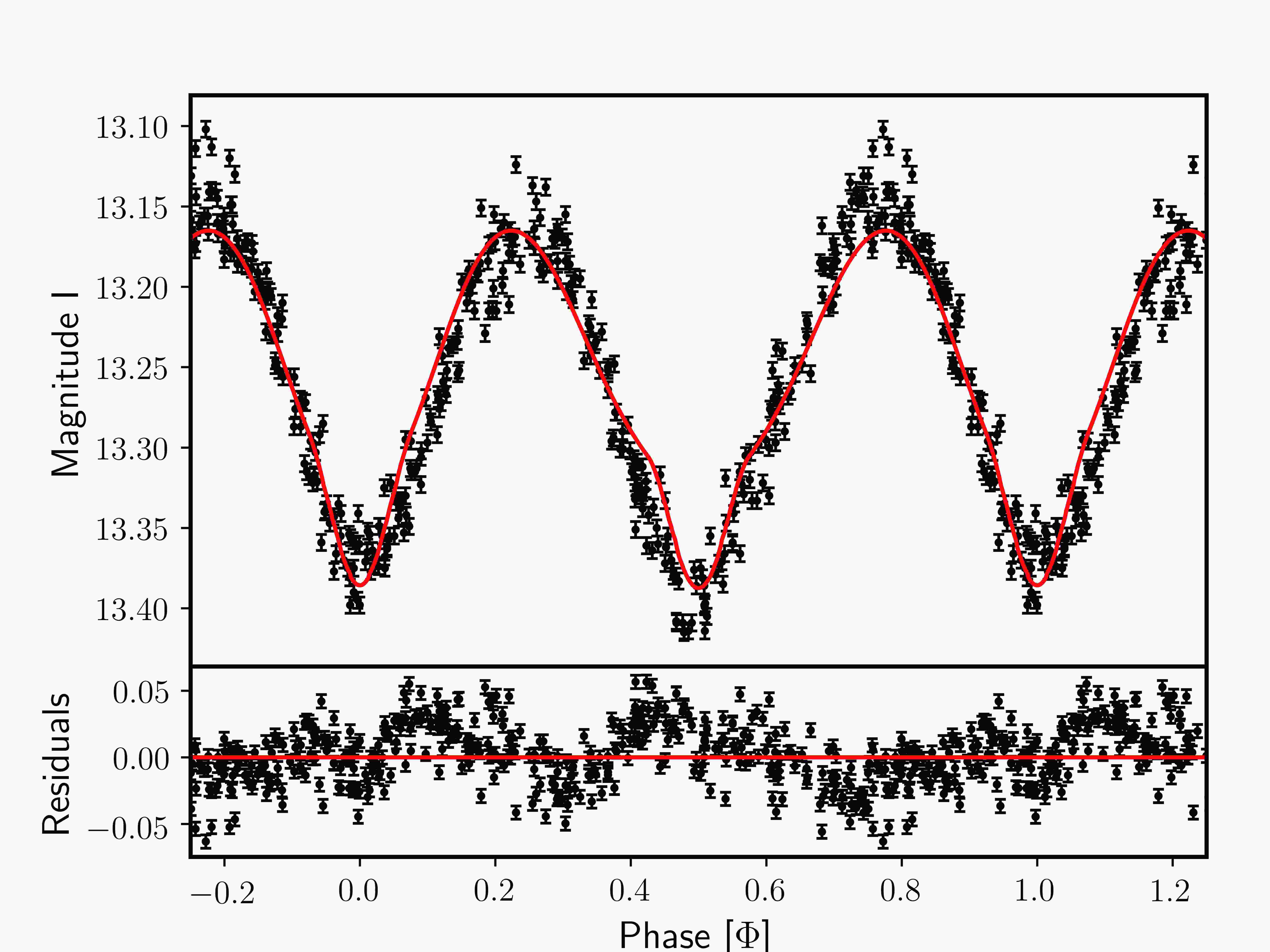}
    \caption{Same as Fig.\,\ref{fig:061}, but for VFTS\,450.}\label{fig:450} 
\end{figure*}

\begin{figure*}[htbp]
  \centering
    \includegraphics[width=9cm, angle=0]{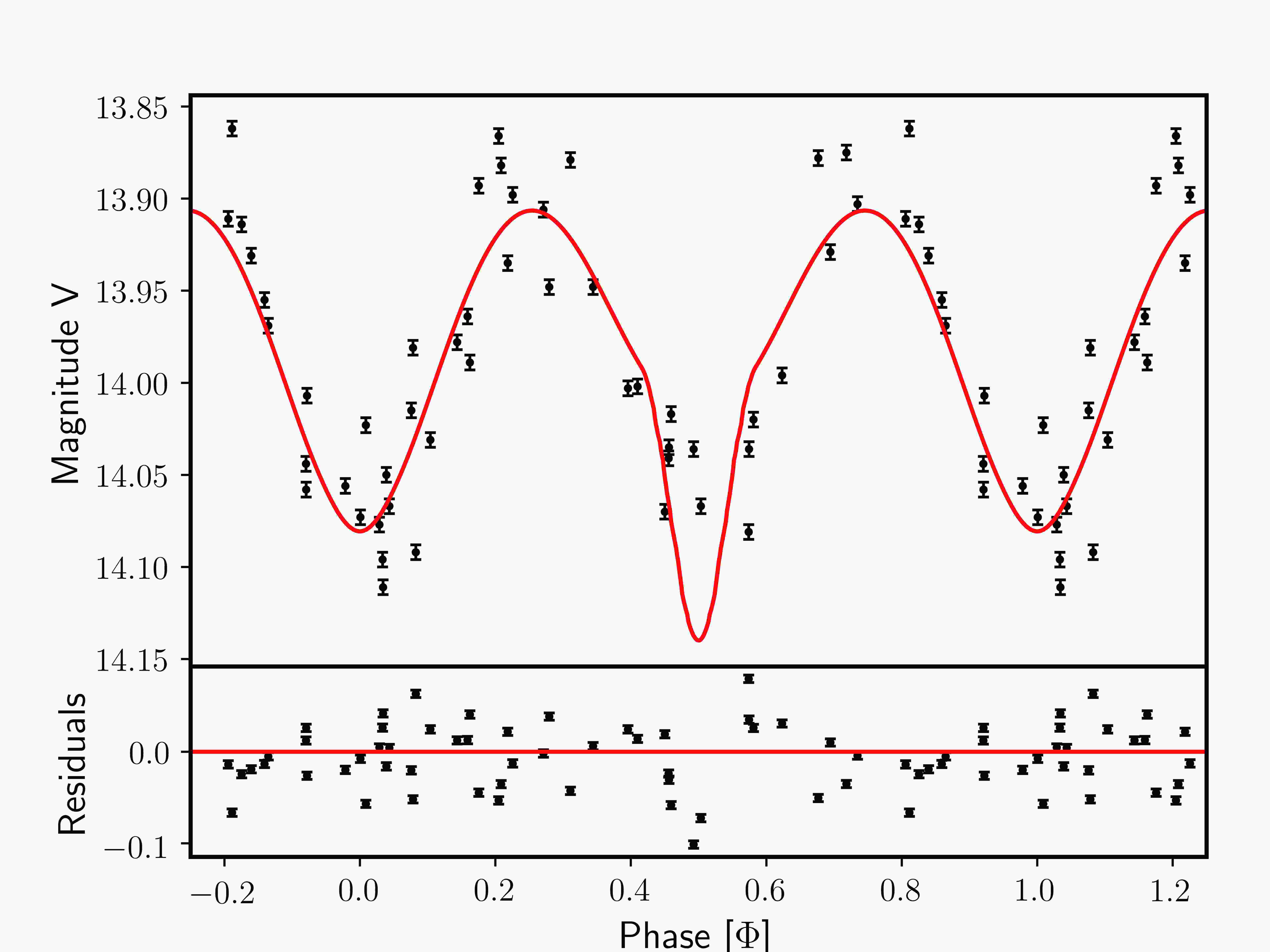}
    \includegraphics[width=9cm, angle=0]{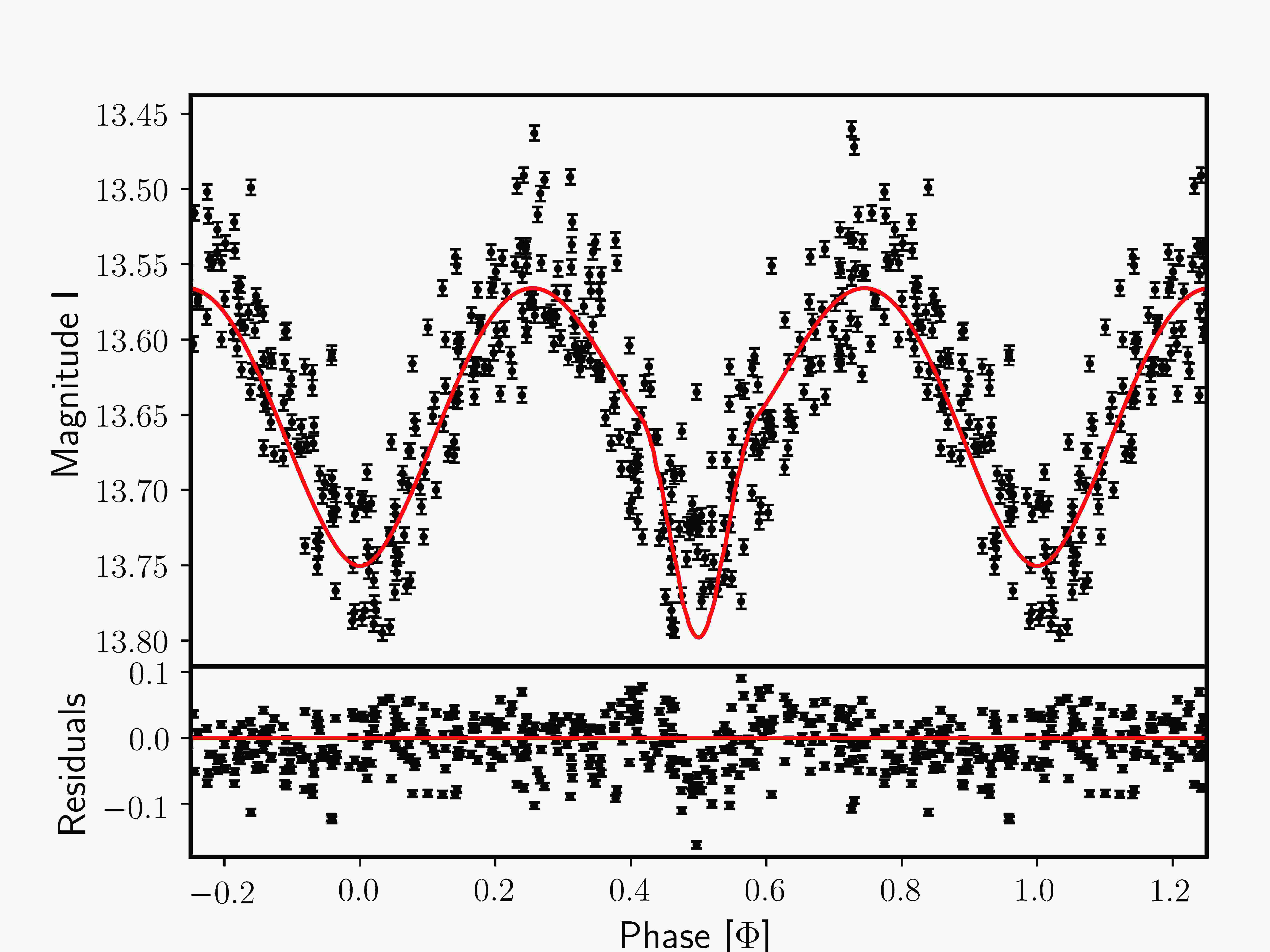}
    \caption{Same as Fig.\,\ref{fig:061}, but for VFTS\,652.}\label{fig:652} 
\end{figure*}

\begin{figure*}[htbp]
  \centering
    \includegraphics[width=9cm, angle=0]{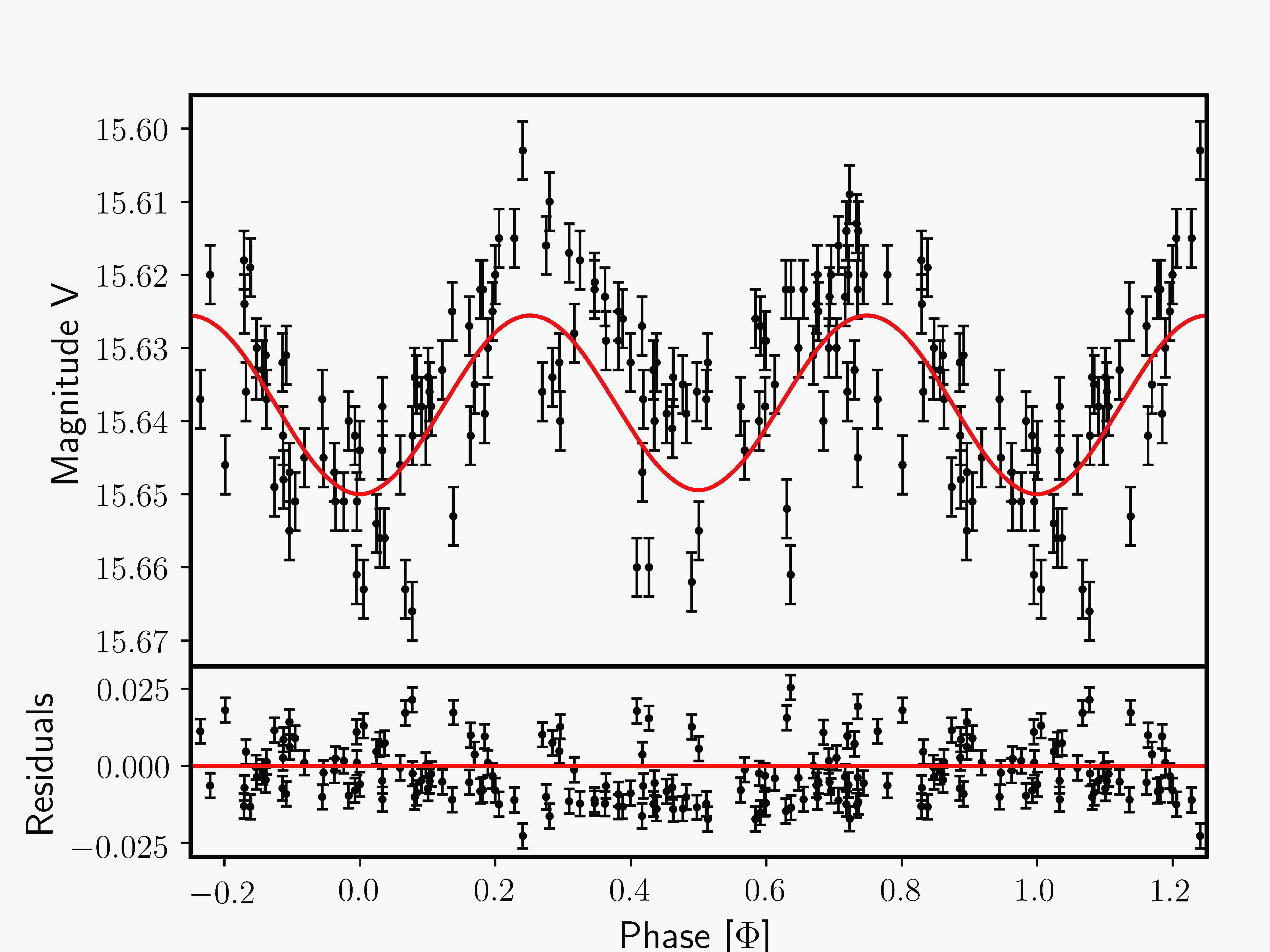}
    \includegraphics[width=9cm, angle=0]{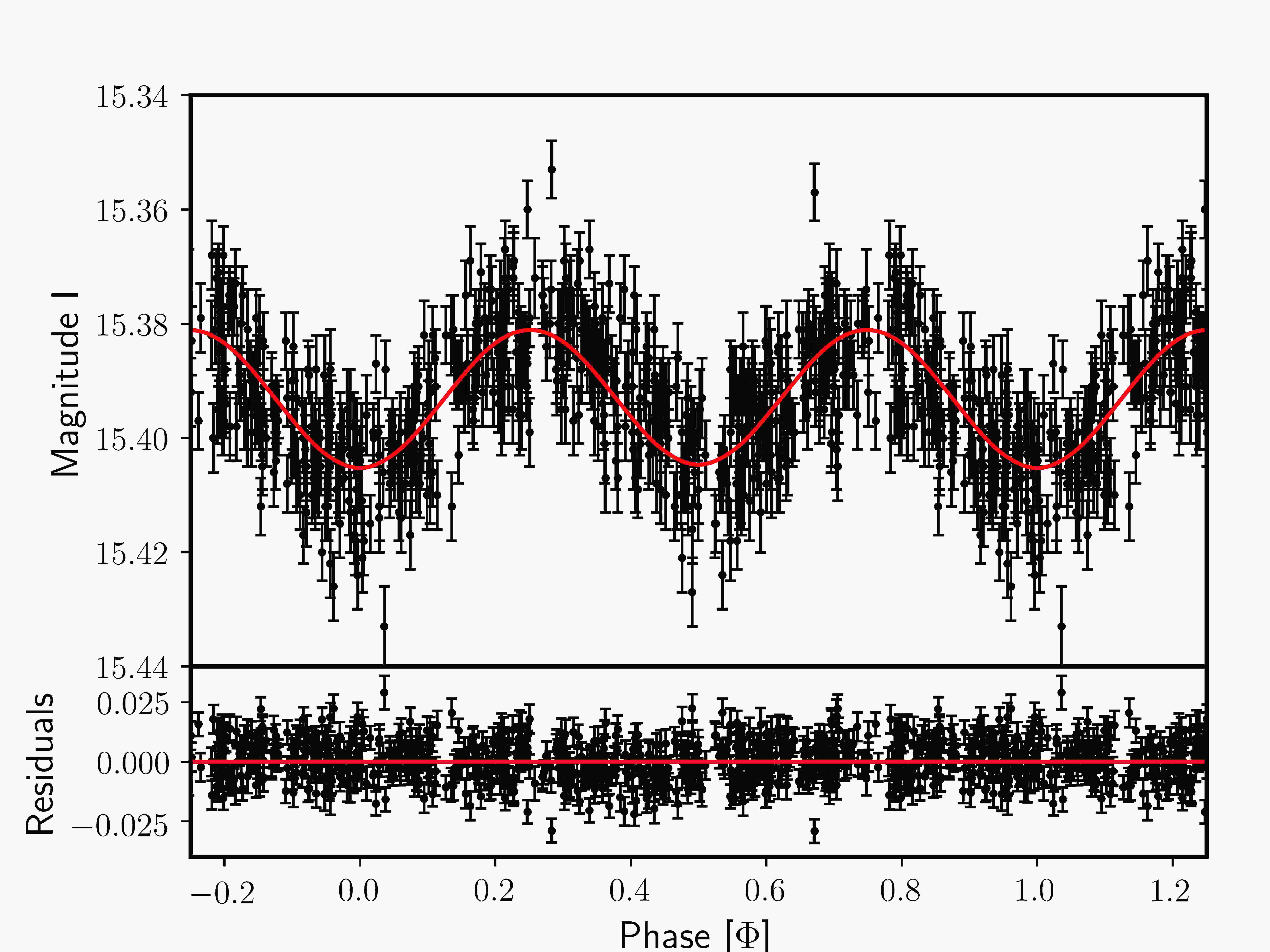}
    \caption{Same as Fig.\,\ref{fig:061}, but for VFTS\,066}\label{fig:066} 
\end{figure*}

\begin{figure*}[htbp]
  \centering
    \includegraphics[width=9cm, angle=0]{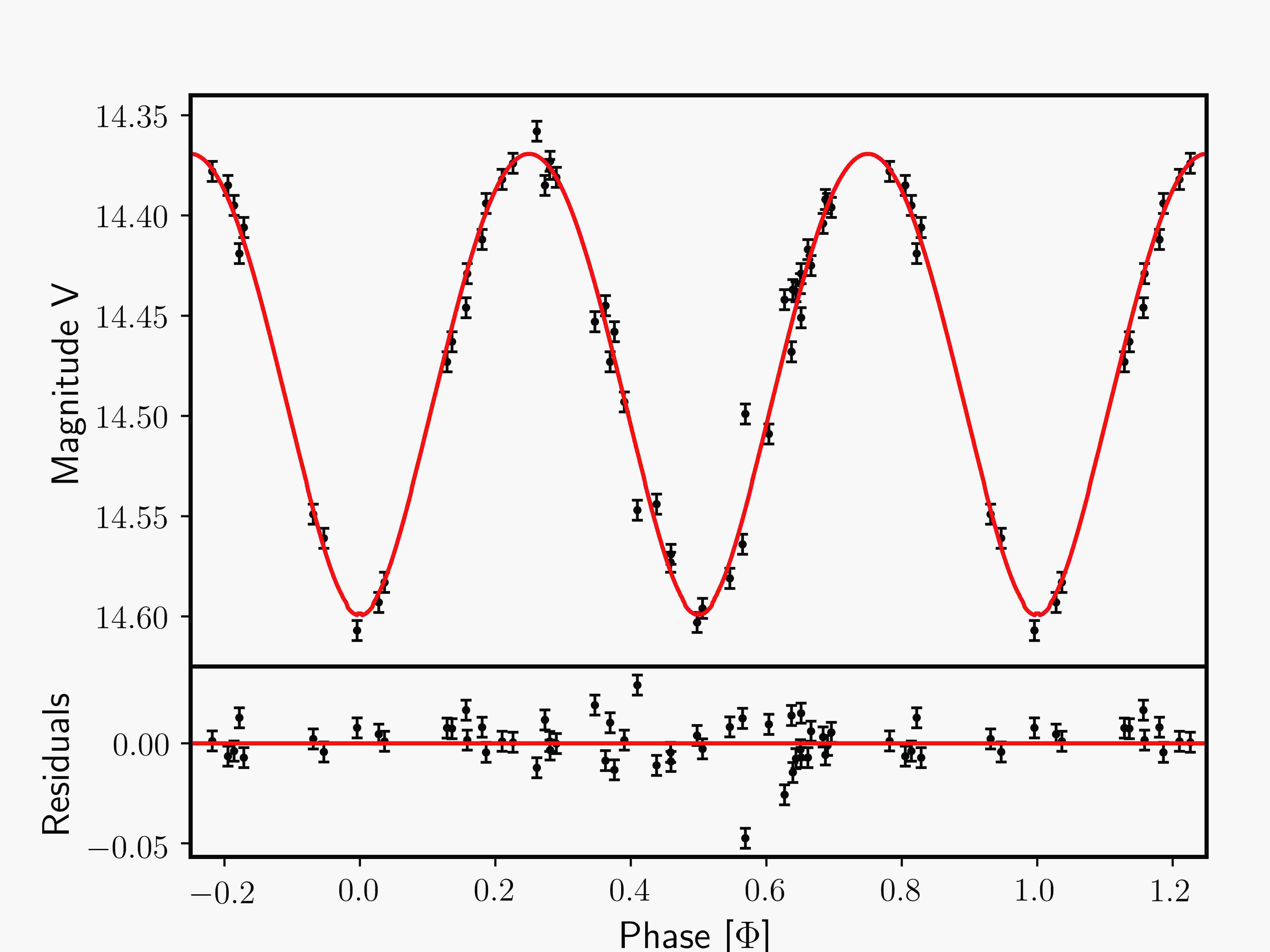}
    \includegraphics[width=9cm, angle=0]{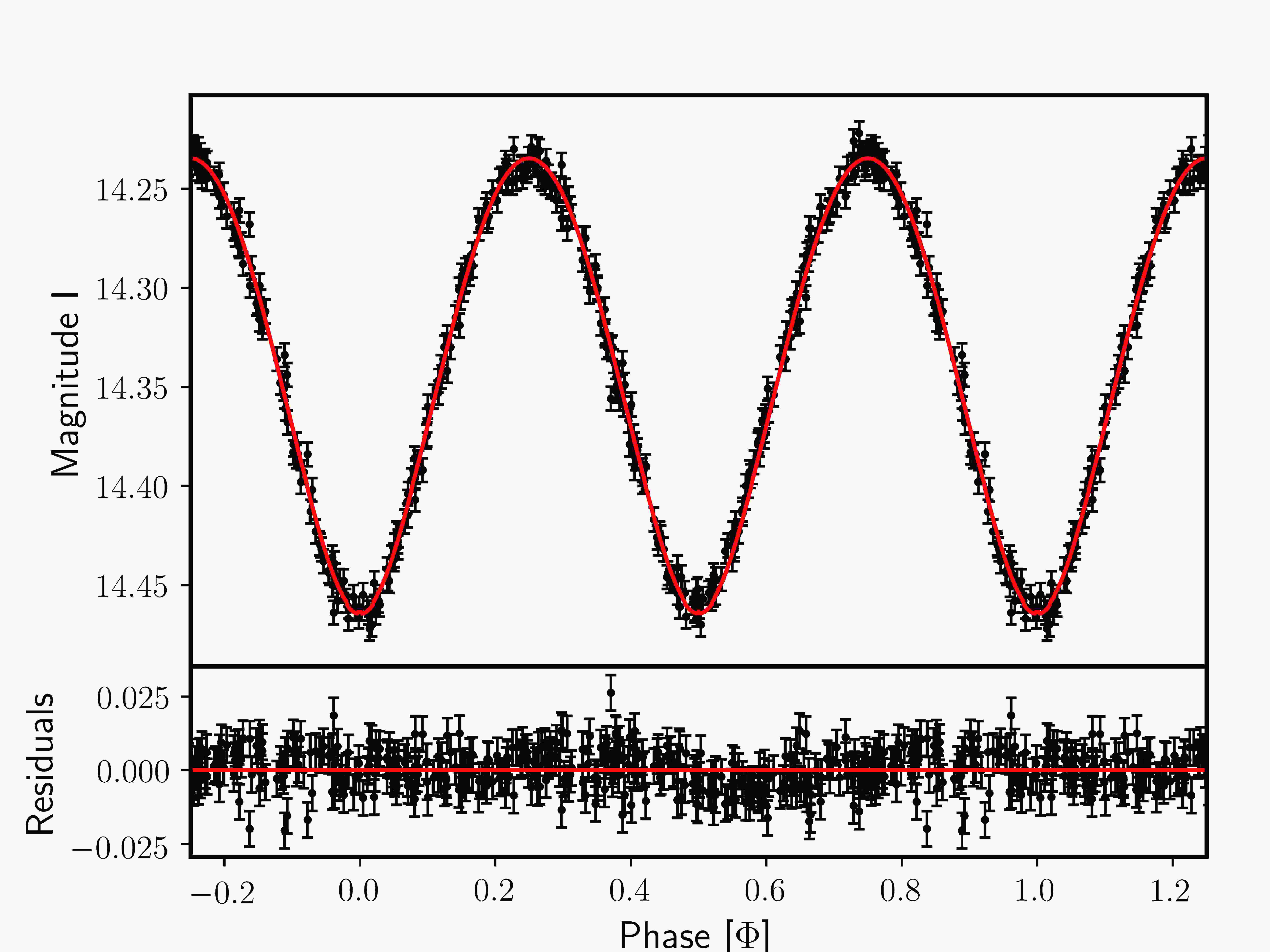}
    \caption{Same as Fig.\,\ref{fig:061}, but for VFTS\,352.}\label{fig:352} 
\end{figure*}

\begin{figure*}[htbp]
  \centering
    \includegraphics[width=9cm, angle=0]{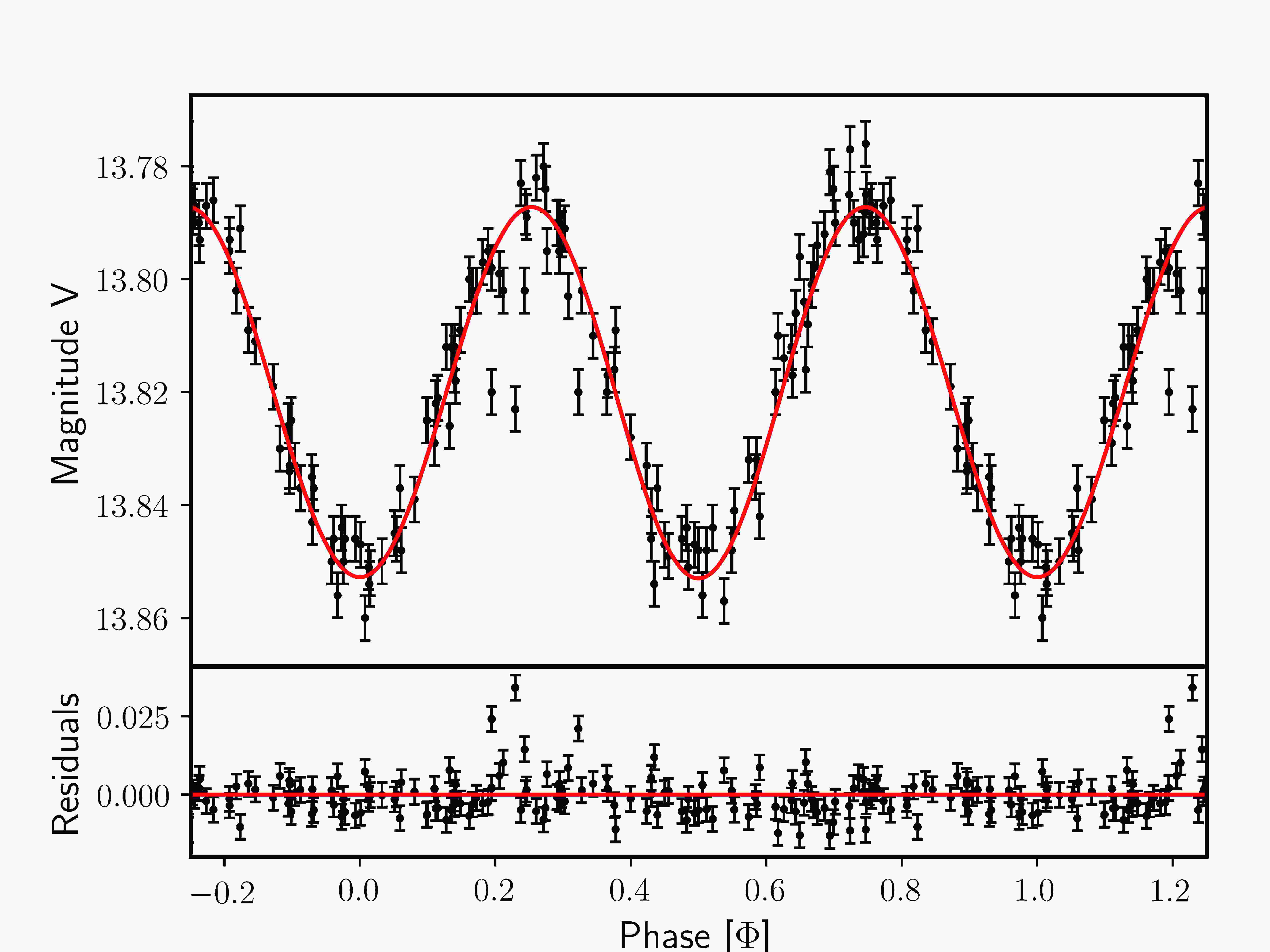}
    \includegraphics[width=9cm, angle=0]{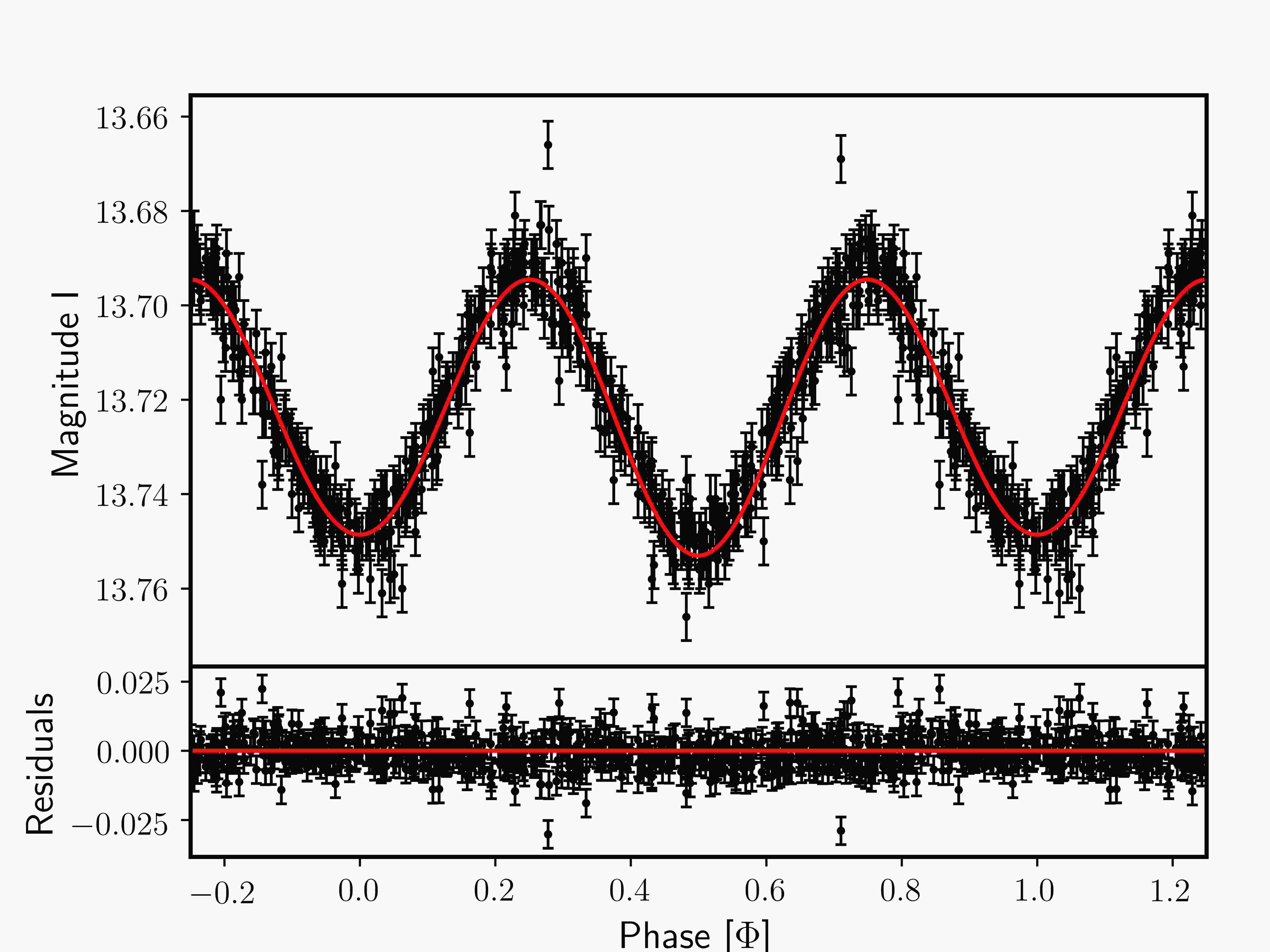}
    \caption{Same as Fig.\,\ref{fig:061}, but for VFTS\,217.}\label{fig:217} 
\end{figure*}

\begin{figure*}[htbp]
  \centering
    \includegraphics[width=9cm, angle=0]{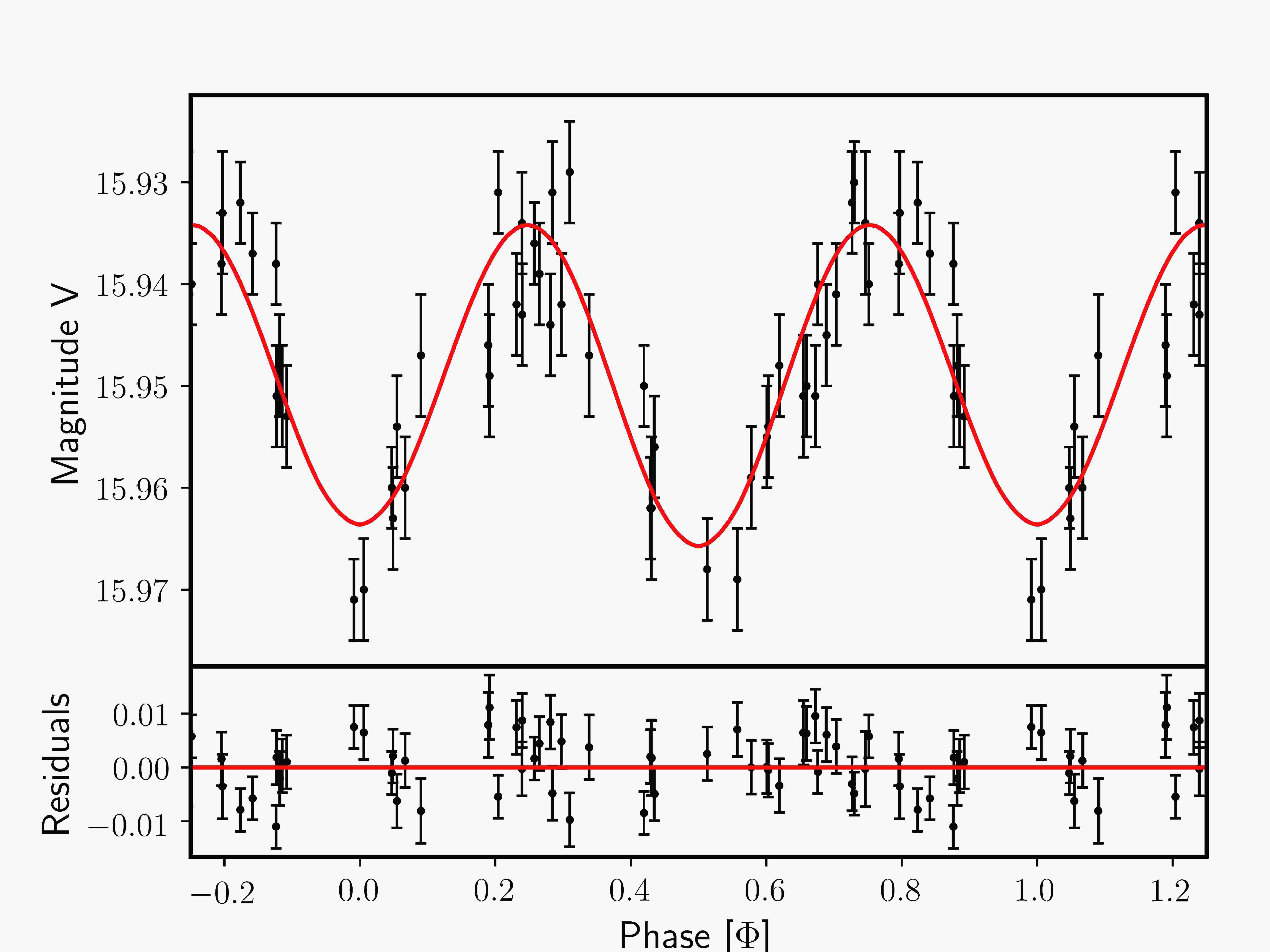}
    \includegraphics[width=9cm, angle=0]{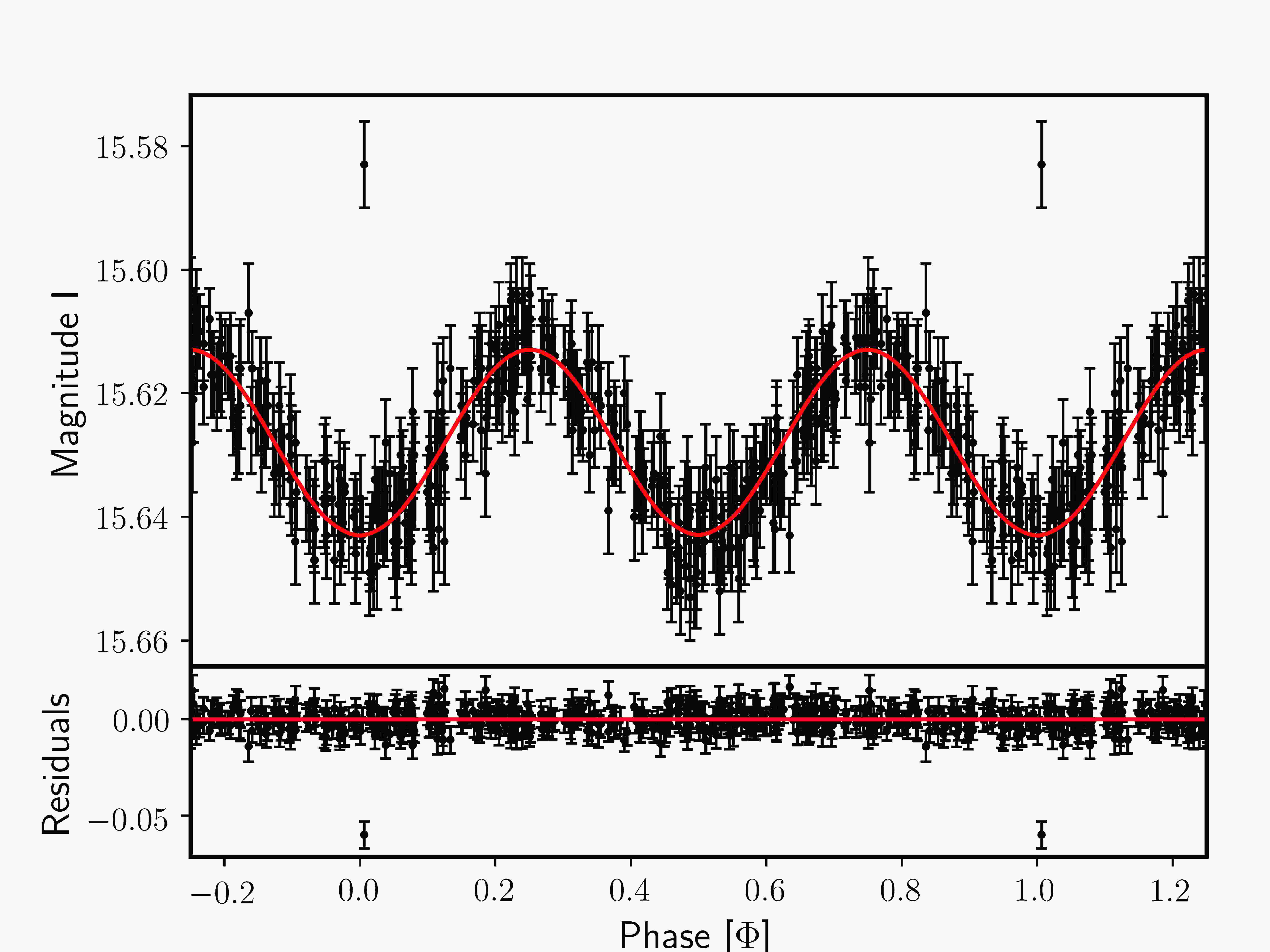}
    \caption{Same as Fig.\,\ref{fig:061}, but for VFTS\,563.}\label{fig:563} 
\end{figure*}

\end{appendix}
\end{document}